\documentclass[11pt,a4paper,english,german]{article}
\usepackage{graphicx}
\usepackage{epsfig}
\usepackage{times}
\usepackage{amssymb}
\usepackage{amsmath}
\usepackage[ruled,vlined]{algorithm2e}

\usepackage[sort&compress,numbers]{natbib} 

\usepackage{dcolumn} 

\begin{document}

\renewcommand{\textfraction}{0.05} 
\renewcommand{\topfraction}{0.95}
\renewcommand{\bottomfraction}{0.95} 
\renewcommand{\floatpagefraction}{0.95}

%%%%%%%%%%%%%%%%%%%%%%%%%%
\newcommand{\mean}[1]{\left\langle #1 \right\rangle}
\newcommand{\abs}[1]{\left| #1 \right|}
\newcommand{\comment}[1]{\vspace{0.5cm}\textit{#1}\vspace{0.5cm}}

\renewcommand{\thefootnote}{ \fnsymbol{footnote} }
%%%%%%%%%%%%%%%%%%%%%%%%%

\title{ADAPTIVE INVESTMENT STRATEGIES FOR PERIODIC ENVIRONMENTS}
\author{ \Large J.-Emeterio Navarro}

\footnotetext{\it{Paper submitted to Advances in Complex Systems (November, 2007)}}

\date{}

\maketitle

\thispagestyle{empty} 

\begin{quote}
  \centering{
    Arbeitsgruppe K\"unstliche Intelligenz\\ Institut f\"ur Informatik\\ Humboldt-Universit\"at zu Berlin\\ Unter den Linden 6\\ 10099 Berlin, Germany}
\end{quote}

\begin{abstract}
  In this paper, we present an adaptive investment strategy for environments with periodic returns on investment. 
  In our approach, we consider an investment model where the agent decides at every time step the proportion of wealth to invest in a risky asset, keeping the rest of the budget in a risk-free asset. 
  Every investment is evaluated in the market via a stylized return on investment function (RoI), which is modeled by a stochastic process with unknown periodicities and levels of noise.
  For comparison reasons, we present two reference strategies which represent the case of agents with zero-knowledge and complete-knowledge of the dynamics of the returns.
  We consider also an investment strategy based on technical analysis to forecast the next return by fitting a trend line to previous received returns. 
  To account for the performance of the different strategies, we perform some computer experiments to calculate the average budget that can be obtained with them over a certain number of time steps.
  To assure for fair comparisons, we first tune the parameters of each strategy. 
  Afterwards, we compare the performance of these strategies for RoIs with different periodicities and levels of noise. 
\end{abstract}

%\keywords{genetic algorithms; portfolio optimization; investment strategies; time series.}

%%%%%%%%%%%%%%%%%%%%%%%%%%%%%%%%%%%%%%%%%%%%%%%%%%%%%%
\section{Introduction}
\label{sec:Introduction}

Finding a proper investment strategy is a problem that has been addressed by many researchers from different areas.
In economy, this problem usually concerns the behavior that an investor should follow in order to maximize the profits under an uncertain environment. 
To this end, researchers usually investigate the relation between methods for optimization under uncertainty, the different preferences of an investor and the amount of information available from the environment.
For this, different measures of risk aversion have been proposed together with the classification of investors by their behavior towards risk (e.g. risk-averse, risk-neutral or risk-seeking behaviors), see \citep{Artzner99}.
Many researchers have been also concerned in finding different manners to control the \emph{risk-exposure}.
Many of the proposed methods are based on decision-making and utility theory and are addressed to scenarios where the investor can choose between investing in a risky or a risk-free asset, see \citep{Kelly56,Kahneman-Tversky79,Kahnemann-Riepe98,Bak-Norrelykke99}.
And other researchers have extended this to the problem of portfolio diversification, where more than one risky asset is considered, see \citep{markowitz52,marsili98,maslov98}.

On the other hand, many researchers have used different machine learning methods to find good investment strategies in different type of stochastic environments.
For example, in \citep{Ismail01} the authors use neural networks to find patterns from financial time series, where the main goal is to find changes in volatility. 
And in \citep{Geibel05}, the authors propose the use of a risk-sensitive reinforcement learning algorithm to find the most proper policy for controlling under constraints and applied it to the control of a feed tank with stochastic inflows.
Other techniques from machine learning that are frequently used for investment decision problems are those based on \emph{evolutionary computation}.
For example, those using \emph{genetic programming} and \emph{genetic algorithms} for portfolio management, inducing rules for bankruptcy prediction, and assigning credit scoring, see \citep{Dawid99}.
Some investment strategies based on genetic programming techniques usually lead to profitable trading strategies, however, they usually find strategies which are difficult to understand and sometimes they cannot be funded \citep{Neely97,schulenburg99evolutionary,schulenburg01strength,Jiang03}.
Even though investment strategies that are based on genetic algorithms may be also difficult to abstract and to explain, we believe that they are more natural and understandable than those using genetic programming techniques \citep{Drake-Marks02}.
However, many of these approaches are applied to environments that are stationary; this means that some of them cannot be directly applied to changing environments. 
In the literature, there are some researches which have investigated the use of genetic algorithms in changing environments \citep{Branke99,Grefenstette92}. 
However, to our knowledge, they have not been applied specifically to the problem of controlling the proportion of investment in periodic environments.

On the other hand, the typical scenario to study investment strategies is to let an agent choose between betting in a lottery or receiving a constant amount of money \citep{Arrow65}.
This simple scenario is usually extended to different type of investment models where investors are commonly referred to as agents and the complexity of the investment models may differ considerably, see \citep{LeBaron00,Farmer01,Lux-Marchesi02}.
In some of these models the amount of money that the agents invest in the market is assumed to be proportional to their budget, this assumption is also called investment fraction or investment proportion. 
Researchers investigate in these models from the optimal investment strategies to the different properties that emerge in the artificial market, see \citep{marsili98,maslov98}.
Interestingly, if the market is simply treated as a random variable and the proportion of investment is fixed to a constant value, then it has been shown that eventually the agent looses all its money in the course of time \citep{Sornette-Cont97,NavarroSchweitzer03}.
In order to avoid bankruptcy, the agent may have an income, \citep{Kesten73, Redner90,Sornette-Cont97}, or a budget-barrier may be assumed \citep{Levy-Solomon96}.
Some researchers have investigated different strategies to control the proportion of investment in this type of models for different scenarios \citep{maslov98,NavarroPhysicaA}.
On the other hand, some other authors use different artificial market models to compare the performance of agents with zero-intelligence and rational agents \citep{Gode-Sunder93,Farmer-Patelli04}.

This paper may also draw interest on the research area of pattern recognition of time series. 
In particular, for the cases when there is no prior knowledge of the existance of a periodic signal or of its characteristics, see \citep{szpiro97PhysRevE.55.2557,alvarez01}.
Note that with some small proper changes on the proposed adaptive algorithm, a useful algorithm could be proposed for the detection and measurement of periodic signal in time series.

In this paper, we propose a new approach based on evolution for finding investment strategies in periodic environments.
This paper is organized as follows. Section~\ref{sec:InvestmentModel} describes the investment model where the agent decides at every time step the percentage of wealth to invest in a risky asset keeping the rest in cash.
Section \ref{subsec:StrategyBasedOnEvolution} presents an adaptive investment strategy based on a \emph{Genetic Algorithm} (GA) for environments with periodic time series.
In Section \ref{sec:ExperimentalResults} we present the results obtained for different computer experiments.
We decided to perform our computer experiments in a controlled scenario where the dynamics of the environment are known.
For that reason, we assume that the risky asset is modeled by a stochastic process with changing periodicity and different levels of noise, i.e. stylized exogenous returns, presented in Section~\ref{subsec:ArtificialReturns}.
For the sake of completeness, we compare the performance of the adaptive strategy proposed in this paper with other investment strategies which are discussed in Section \ref{sec:ReferenceStrategies}.
For our comparison, we include two reference strategies which represent the agent with both zero-knowledge and complete-knowledge of the dynamics of the RoI.
We also include an investment strategy based on technical analysis that basically forecasts the next return by fitting a trend line to the previous received returns. 
To account for the performance of the different strategies, we perform some computer experiments to calculate the average budget that can be obtained with them over a certain number of time steps.
The experiments to account for the performance of the investment strategies are divided into two sections.
First, in Section \ref{subsec:RoIWithFixedPeriodicity} we consider a stationary environment, i.e returns with fixed periodicity, and in Section \ref{subsec:RoIWithChangingPeriodicity} we consider a non-stationary environment, i.e return with changing periodicity.
To assure for fair comparisons, we first tune the parameters of each strategy and afterwards, we compare their performance for returns with different levels of noise. 

%%%%%%%%%%%%%%%%%%%%%%%%%%%%%%%%%%%%%%%%%%%%%%%%%%%%%%
\section{Investment Model}
\label{sec:InvestmentModel}

We consider an investment model \citep{NavarroSchweitzer03, NavarroPhysicaA} where an agent is characterized by two individual variables:
(i) its \emph{budget} $x(t)$, i.e. its wealth and  
(ii) its \emph{investment proportion} $q(t)$, i.e. its attitude towards risk in a market. 
The budget, $x(t)$, changes in the course of time $t$ by means of the following dynamic:
\begin{equation}
  \label{eq:wealth}
  x(t+1)=x(t)\,\Big[1+r(t)\,q(t)\Big]
\end{equation}

More in detail, this means that the agent at time $t$ invests a portion $q(t)x(t)$ of its total budget. 
And this investment yields a gain or loss on the market, expressed by $r(t)$, the return on investment, \emph{RoI}. 

Some authors assume that returns are obtained by means of continuous double auction mechanisms \citep{LeBaron00,Lux-Marchesi02}, however, in our approach, we rather consider that the returns are not being influenced by agent's actions.
In other words, we assume that the agent has a small budget and its actions do not affect the evolution of the returns.
Later on, we present more in detail the dynamic for the returns with seasonal market changes, Eq.~(\ref{eq:roi_sin}).

The behavior of the agent in this environment is expressed in terms of its investment proportion, $q(t)$, which corresponds to the percentage or portion of agent's budget that is susceptible to win or lose, i.e. the agent's attitude towards risk. 
 We assume that the agent's investment proportion may change for example, dependent on the agent's predictions or assumptions about the market dynamics.

Since $q(t)$ always represents a portion of the total budget $x(t)$, it is bound to a minimum value of zero and a maximum value of one, i.e. $q(t)\in[0,1]$. 
This means that an agent with $q(t)=0$ decides at time step $t$ to perform no investments at all, whereas an agent with $q(t)=1$, is investing at time step $t$ all its capital.
For the sake of completeness, we assume that the minimal and maximal investment-proportions are described by $q_{\mathrm{\min}}$ and $q_{\mathrm{\max}}$, respectively.

Thus, in this paper we present an adaptive investment strategy, expressed by a method to find the most proper $q(t)$ and we focus on the performance of this investment strategy in periodic returns.
Of course the agent may have some bounded memory about past RoI that could be used for predictions of future RoI. 
And as mentioned above, we assume a simple dynamic for the returns allowing us to focus in the feedback of these market returns on the  investment strategy (and not on the feedback of the strategies on the market).

Last but not least, we assume that the agent invests independently in the  market, i.e. there is no direct interaction with other agents.

\section{Adaptive Investment Strategy}
\label{subsec:StrategyBasedOnEvolution}

In this section, we present an adaptive investment strategy based on a \emph{Genetic Algorithm} for controlling proportions of investment in periodic environments.
For simplicity, we call this strategy \emph{Genetic Algorithm for Changing Environments} (GACE).

\emph{Genetic algorithms} (GA) are stochastic search algorithms based on evolution that explore progressively from a large number of possible solutions finding after some generations the best solution for the problem.  
Inspired by natural selection, these  powerful techniques are  based on  some defined  evolution operators, like selection, crossover and mutation \citep{Holland75,Goldberg89,Forrest96,Michalewicz99}.

In our approach, we consider that an agent uses a GA to find the most proper set of investment proportions for every time step.
For this, we show on the following the specifications for the GA.

\subsection{Encoding Scheme}

A population  of chromosomes $j=1,...,C$, where each  chromosome   $j$ has  an   array   of  genes,   $g_{jk}$, where $k=0,...,G_{j}-1$, and $G_{j}$ is the length of the chromosome $j$.
The length of a chromosome is assumed to be in the range $G_j\in(1,G_{\mathrm{\max}})$, where $G_{\mathrm{\max}}$ is a parameter that specifies the maximal allowed number of genes in a chromosome.
The values of the  genes could be binary, but for programming reasons we use real values, see \citep{Michalewicz99}.   
Moreover, each chromosome $j$ represents a  \emph{set of  possible strategies}  of an  agent, i.e. each $g_{jk}$ corresponds to an investment proportion.

\subsection{Fitness Evaluation}

Each chromosome $j$ is evaluated after a given number of time steps by a \emph{fitness function}, $f_{j}(\tau)$, which is defined as follows:
\begin{equation}
  \label{eq:fitness}
  f_{j}(\tau)=\sum_{k=0}^{G_{j}-1} r(t)\,g_{jk} \;; \quad k \equiv t \bmod G_{j},
\end{equation}
where $\tau$ is a further time scale in terms of generations.
When a generation is completed, the chromosomes' population is replaced by  a  new population  of  better  fitting  chromosomes  with the  same  population  size $C$. 

As you can see, every $g_{jk}$ is multiplied by a different value of $r(t)$ in the course of time.
Since the  fitness of a  chromosome tends to  be maximized, negative $r(t)$ should lead to small values of $g_{jk}$, i.e. small investment proportions. 
On the other hand positive $r(t)$ should lead to larger values of $g_{jk}$, i.e. large investment proportions.
Because of this, we consider the product of $r(t)g_{jk}$ as a performance measure, which is in accordance with our investment model, Eq.~(\ref{eq:wealth}).
Noteworthy, in this approach the GA tries to find the chromosomes leading to larger profits.
Another different approach would be to implement a GA to find the chromosomes that minimize the loss, in which case, we would have a different fitness function.
Also note that chartists usually study the past movement of stock prices; however, this approach differs from ours in the fact that we treat directly returns on investment and not price movements.

\subsection{Selection of a New Population}
\label{subsec:selection}

If we assume that chromosomes have fixed length, $G_{j}=G_{\mathrm{\max}}, \forall j$, then the most proper number of time steps, $t_{\mathrm{eval}}$, that have to elapse in order to evaluate all chromosomes' genes is $t_{\mathrm{eval}}=G_{\mathrm{\max}}$.
In other words, the number of time steps needed to evaluate the population is equal to the fixed length of the chromosomes.

Moreover, it can be shown that the population converges faster towards optimal investment proportions if the length of the chromosomes is equal to the periodicity of the returns, $G_{\mathrm{\max}}=T$.

However, this previous assumption corresponds to the ideal case where the agent knows a priori the periodicity of the returns and sets the length of all chromosomes to the value of the periodicity, hence the agent selects a new population after all genes of all chromosomes are being evaluated.
Thus, if the chromosomes have different length the question now is the following: 
\emph{After how many time steps, $t_{\mathrm{eval}}$, a new generation of chromosomes should be obtained?} 
In the following, we propose different approaches to answer this question.

\subsubsection{Time steps for evaluation}
\label{sec:TimeForEvaluation}

Different approaches can be proposed to determine the number of time steps $t_{\mathrm{eval}}$ that should elapse to select a new generation of chromosomes.
As mentioned above, the simplest approach, called \emph{GMaximum}, is to select a new population after a fixed number of time steps.
If $G_{\mathrm{\max}}$ is equal to the maximal length of the chromosomes, $t_{\mathrm{eval}}=G_{\mathrm{\max}}$, then all chromosomes' genes in the population will be evaluated.
However, such an approach leads to slow convergence of the population.
A different approach may be to choose the number of time steps for evaluation accordingly to the length of the best chromosome in the population. 
This approach is called \emph{GBestSelected}, and it can be expressed mathematically as follows: $t_{\mathrm{eval}}=G_{l}$ with $l=\mathrm{arg}\; \mathrm{max}_{j=1..,C}\; f_{j}(t'_{\mathrm{eval}}))$., where $t'_{\mathrm{eval}}$ is the number of time steps that the population has been evaluated.
This approach leads to a faster convergence of the population than when using GMaximum; however, if the length of the best chromosome in the previous generation happen to be very large, this would lead to a larger number of time steps using only this strategy.
This would be disadvantageous for the agent if the strategy actually leads to looses instead of profits for the current returns.
A better approach is to choose the the number of time steps needed for evaluation according to the length of the best chromosome at every time step $t$. 
This approach is called \emph{GBestCurrent}, and can be expressed mathematically as follows: $t_{\mathrm{eval}}=G_{l}$ with $l=\mathrm{arg}\; \mathrm{max}_{j=1,..,C}\; f_{j}(t))$. 

Note that the last two approaches have the disadvantage that they do not assure that all genes of all chromosomes are being evaluated; however, from our point of view, good chromosomes would lead to larger fitness than bad ones from the very beginning of the evaluation.
It can happen that by coincidence the cycle of the returns match exactly a small number of good genes in bad chromosomes; however, on the long run only the good chromosomes would subsist.
Unless otherwise indicated, we assume on the following that the approach GBestCurrent is being used for the evaluation of the population.

\subsubsection{Elitist and Tournament Selection}

Once the time has come to select a new population, the question is: \emph{how to determine a new  population?} 

After calculating the fitness of each  chromosome  according to  Eq.~(\ref{eq:fitness}),  we first find the  best chromosomes  from  the  old  population  by applying  elitist  and  tournament selection of size two.  
Elitist selection considers the best $s$ percentage of the population  which is found by  ranking the chromosomes  according to their fitness.   
These   best  chromosomes  are  directly  transferred   to  the  new population.
Afterwards, a tournament  selection is done by randomly choosing  two pairs  of two chromosomes  from the old  population and then selecting  from each  pair the  one with the  higher fitness.   
These two chromosomes are  not simply  transferred to the  new population, but  undergo a transformation  based on  the  genetic operators' crossover and mutation.

\subsubsection{Crossover and Mutation Operators}

Once two chromosomes have been selected by means of the tournament selection, a simple crossover operator would be one that exchanges genetic information between the two chromosomes, whatever their sizes, by finding the cross-point with respect to the size of the shortest chromosome.
More in detail, this is done by selecting randomly from the shortest chromosome the cross point or cut point, $c_p$, and with probability $p=0.5$ to exchange the genetic material above or beyond this cross point in the shortest chromosome with its counterpart in the largest chromosome.
However, those genes in the largest chromosome beyond the length of the shortest chromosome would be disregarded.

The limitations of conventional crossover in GA with variable length has already been addressed by some authors \citep{Harvey92}, where neural networks or hierarchical tree-structures are used to determine which genes should be exchanged between the chromosomes. 
However, for the purpose of this paper and for the sake of simplicity, we propose a modification of the standard GA crossover operator that better suits our demands.

Thus, we propose the use of a crossover operator called \emph{Proportional Exchange Crossover} (PEC) operator, which basically shrinks or stretches the genetic information between the pair of chromosomes proportionally to their length. 
Basically, the crossover operator PEC first randomly selects the range of genetic information to be exchanged between two chromosomes and contracts(extends) the genetic information from the largest(shortest) to the shortest(largest) chromosome, respectively.

More in detail, the Algorithm \ref{algo:pec} shows the PEC algorithm for all pair of parent-chromosomes being selected via tournament selection.
Note that a chromosome $j$ is saved in an array with indexes in the range 0 to $G_{j}-1$.

%%%new using algorithm package

\incmargin{0em}
\linesnumbered
\begin{algorithm}[]
\dontprintsemicolon
\SetLine
\caption{Proportional Exchange Crossover (PEC) operator}
\label{algo:pec}
\ForEach{pair of parent-chromosomes}{
  determine the size $G_{s}$ of the shortest parent-chromosome $pa_{s}$\;
  find the cross-point, $cp_{s}\in \mathbb{Z}$, for the shortest parent-chromosome: $cp_{s} \sim U(0,G_{s}-1)$\;
  determine the size $G_{l}$ of the largest parent-chromosome $pa_{l}$\;
  find, the cross-point $cp_{l} \in \mathbb{Z}$ for the largest parent-chromosome: $cp_{l}=\frac{G_{l}\; cp_{s}}{G_{s}}$ \; \label{line:cpl}
  determine the proportion $R \in \mathbb{Z}$ between the two chromosomes' sizes: $R=\frac{cp_{l}}{cp_{s}}$ \;
  create two arrays, $ch_{s}$ and $ch_{l}$, for the short and large children-chromosomes\;
  with equal probability choose the side for the crossover operation\;
  \eIf{crossover on the left side}{
    extend the genetic material from $pa_{s}$ and copy it to $ch_{l}$ as follows:\;
    \For{$m=0$ to $cp_{s}-1$}{ 
      \For{$n=0$ to $R-1$}{
        $ch_{l}[m*R+n] \leftarrow \mathrm{extend}(pa_{s},m,R)$  \; \label{line:Extend}
      }
    }
    contract the genetic material from $pa_{l}$ and copy it to $ch_{s}$ as follows: \;
    \ForEach{$m=0$ to $cp_{s}-1$}{
      $ch_s[m] \leftarrow \mathrm{contract}(pa_{l},m,R)$ \;      \label{line:Contract}
    }
  }{ %else crossover on the right side
    extend as in line \ref{line:Extend} but for the range $m=cp_{s}$ to $G_{s}-1$.\;
    contract as in line \ref{line:Contract} but for the range $m=cp_{s}$ to $G_{s}-1$.\;
  }
  copy directly the rest genetic material from the parents to the children chromosomes.
}%end for
\end{algorithm}

Note, that different functions could be considered for the transformation of the genetic material between chromosomes with different length.
For simplicity, we consider in our computer experiments a PEC operator based on averaging and copying the genetic material of the parent-chromosomes.
This means that in Algorithm \ref{algo:pec}, we consider for our implementation of GACE: in line 13 the function $extend(pa,m,R)=pa[m]$, which simple copies the genes from the short parent-chromosome to the large child-chromosome; and in line 18 the function $contract(pa,m,R)=1/R\; \sum_{i=m}^{m+R}pa[i]$, which performs an average over the genetic material.
A more interesting option for this transformations could be based on the dynamic time warping algorithm \citep{sankoff83} which is usually used for the calculation of the similarity between two signals. 
With some modifications, this algorithm could be used to stretch or to shrink the genetic material proportionally to the original material; however, this is far from the scope of this paper.

\begin{figure}[th]
    \centerline{\psfig{file=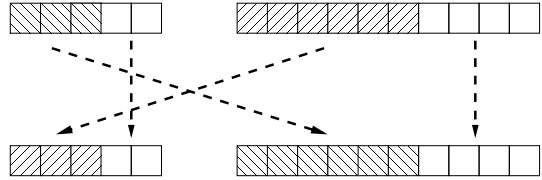,width=7cm}}
    \vspace*{8pt}
    \caption{Example of the \emph{Proportional-sized Exchange Crossover} (PEC) operator. With probability $p=0.5$ the left side of the cut point is selected for exchange.}
    \label{fig:pec}
\end{figure}

To illustrate how the PEC operator works, we show in Figure~\ref{fig:pec} a pictorial representation of PEC applied to the left side of the cross-point.
In this example the cross-point of the shortest chromosome is $cp_{s}=3$.
Consequently, using line 5 in Algorithm~\ref{algo:pec}, we find that the cross-point for the largest chromosome corresponds to $cp_{l}=6$.
In this example the genes to the left of the shortest ``parent'' chromosome are generalized into the largest ``child'' chromosome, whereas the genes to the right of the cross-point are directly copied into the shortest ``child'' chromosome.
The same occurs for the genes in the largest ``parent-chromosome'' with the main difference that the value of the genes to the left are averaged and not generalized.
If the right side of the crossover is selected, we determine in the same manner the cross-points in the ``parent'' chromosomes and we obtain the gene values for the ``child'' chromosomes.

Now, to make sure that a population with chromosomes of diverse lengths is present, we introduce a mutation operator for the length of the chromosome, $G_j$. 
For this, a new length is drawn randomly and the genetic information of the chromosome is proportionally scaled to the new length.
In other words, this operator mutates the length of the chromosome $G_j$ with probability $p_l$ leading to a new enlarged or stretched chromosome. 
The algorithm used for the mutation of the length of the chromosome is based on the same principle as the PEC operator.

Thus, the combination of the PEC operator and the mutation in the chromosomes' length may help to determine the optimal investment proportions and the periodicity (or patterns) of the returns, respectively.

After the crossover and length-mutation operators are applied, the typical gene-mutation operator is applied. 
This means that with a given mutation probability $p_{m}\in  U(0,1)$, a gene is to be mutated by replacing its value by a random number from a uniform distribution  $U(q_{\mathrm{\min}},q_{\mathrm{\max}})$.  

Summing it up, given a population with $C$ chromosomes, to obtain a new generation of chromosomes one needs to do the following: 

\renewcommand{\labelenumi}{\Roman{enumi}}
\begin{enumerate}
\item apply the elitist operator to select the best $s$ percent of the population which are directly included in the new population.
\item  the tournament selection operator is applied to the current population to select two ``well-fitted'' parents.
\item  with probability $p_c$, the PEC crossover operator is applied to the two selected parent-chromosomes yielding  two children-chromosomes.
\item  with probability $p_{l}$, we apply the length-mutation operator to the two children to ensure length diversity in the new population.
\item  with probability $p_{m}$, the gene-mutation operator is applied to the two children which are then included in the new population.
\item  and finally, steps II to V are repeated until the new population has the same number of chromosomes as the original population.
\end{enumerate}

\subsubsection{Strategy Selection and Initialization}

Once a new population has been obtained, we need to answer the following question: \emph{how does the agent update its actual investment proportion, $q(t)$?}

For every new generation, the  agent takes the set of  strategies $g_{jk}$  from the chromosome $j$ with the largest  fitness in  the previous generation. 
\begin{equation}
  \label{eq:ga-q} 
  q_{i}(t) = g_{lk}
  \quad \mathrm{with}\; l = \arg\,\max\nolimits_{j=1,...,C}\, f_{j};
  \; k \equiv t \bmod G_l
\end{equation}

For the initialization, each $g_{jk}$ is assigned  a random value drawn from a Uniform distribution: $g_{jk}\sim U(q_{\mathrm{\min}},q_{\mathrm{\max}})$.
And the length of the chromosomes can be set initially to a fixed number of genes or it can be determined randomly. 
For the latter, each $G_{j}$ is initialized with an integer random value drawn from a Uniform distribution, where $G_{\mathrm{\max}}$ is the maximal allowed chromosome length.

%%%%%%%%%%%%%%%%%%%%%%%%%%%%%%%%%%%%%%%%%%%%%%%%%%%%%%
\section{Experimental Results}
\label{sec:ExperimentalResults}

In this section we systematically analyze the performance of the strategies presented previously.
For this, we present in Sec.~\ref{subsec:ArtificialReturns} the environment for the agent, i.e. the returns, and for each environment we first investigate the parameter tuning of the Genetic Algorithm by means of increasing systematically the complexity of the operators and the environment and we finally compare the performance of the investment strategies presented in this paper.

\subsection{Artificial Returns}
\label{subsec:ArtificialReturns}

First, we consider artificially generated returns, which are driven by the following dynamics:
\begin{equation}
  \label{eq:roi_sin} 
  r(t)=(1-\sigma) \sin \left( \frac{2\pi}{T(t)}\,t \right) + \sigma \xi,
\end{equation}
where the amplitude of the sinusoidal function depends on the amplitude noise level $\sigma \in (0, 1)$, and $\xi$ corresponds to a random number drawn from a Uniform distribution, $\xi \in U(-1,1)$. 
The periodicity of the returns depends on the current time step and would be present for a number of $t'$ time steps, for initial $t'=0$, we have:
\begin{eqnarray}
  \label{eq:StochasticT}
  \textrm{if $t<t'$} &\Rightarrow&  T(t) = T(t-1) \\
  \textrm{else} & \Rightarrow & 
  \begin{cases}
    T(t) = \tilde{T} \\
    t'= t+\tilde{t},
  \end{cases}
\end{eqnarray}
where both $\tilde{T}$ and $\tilde{t}$ are random numbers drawn from the Uniform distributions $U(0,T_{\mathrm{\max}})$ and $U(0,t_{\mathrm{\max}})$, respectively.

Thus, $\sigma$ accounts for the fluctuations in the market dynamics on the amplitude of the RoI; $T_{\mathrm{\max}}$ accounts for the largest possible periodicity and $t_{\mathrm{\max}}$ accounts for the maximal number of time steps a periodicity can elapse.
 
\begin{figure}[th]
  \centerline{
    \psfig{file=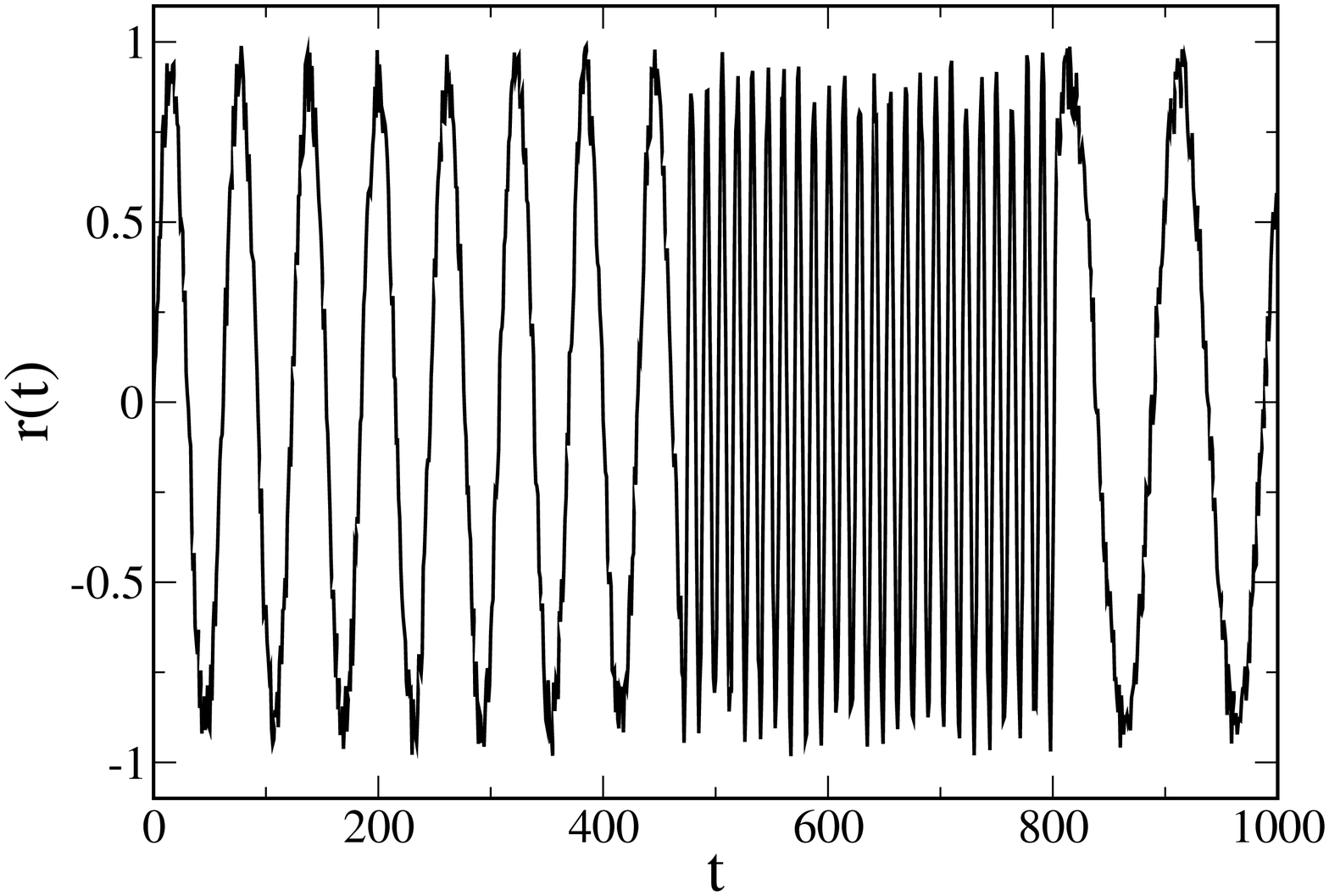,width=8.5cm}
    \psfig{file=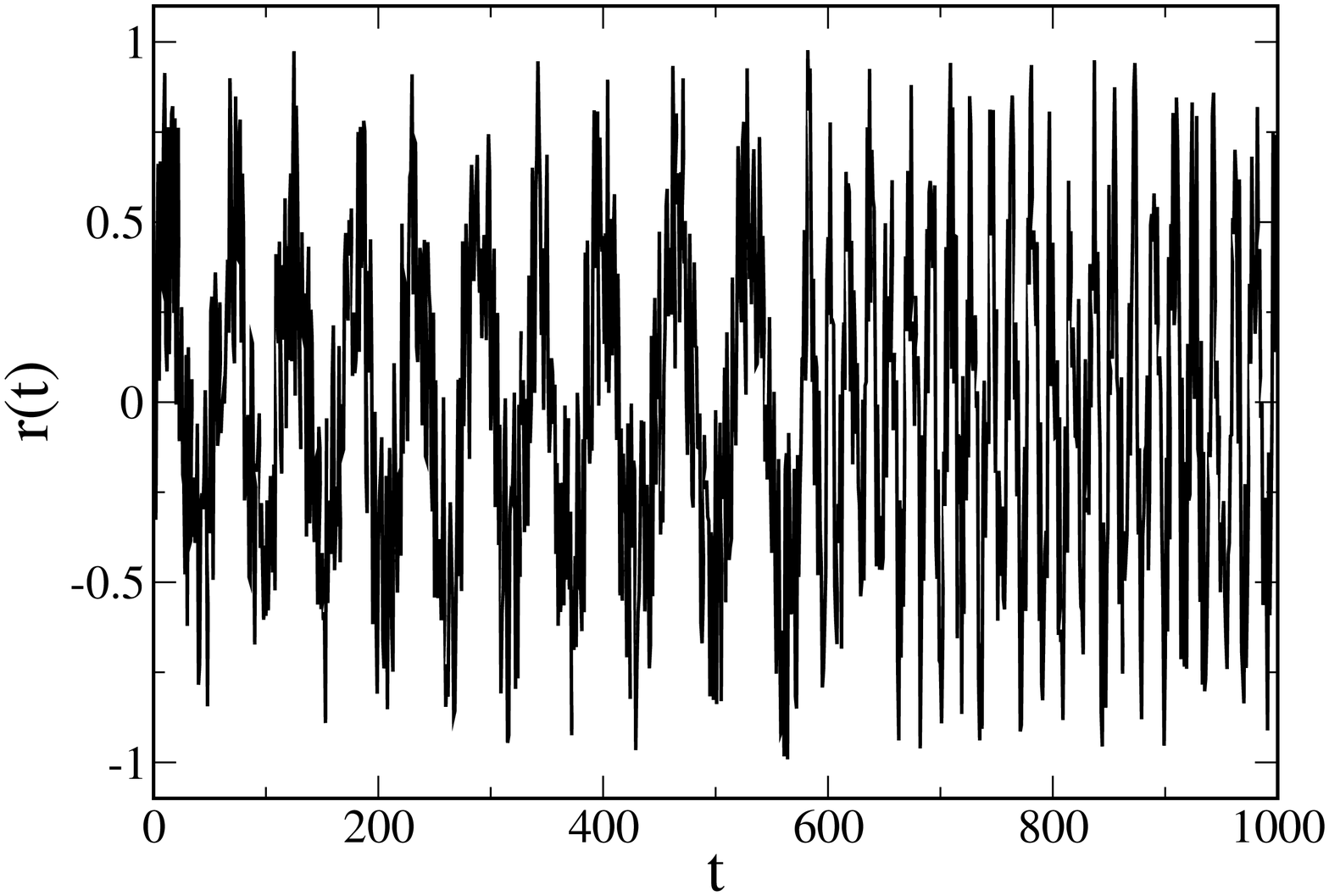,width=8.5cm}}
  \vspace*{8pt}
  \caption{Periodic RoI, $r(t)$, Eq.~(\ref{eq:roi_sin}) for different amplitude fluctuations: (left) $\sigma=0.1$, and (right) $\sigma=0.5$. Further parameters: $T_{\mathrm{\max}}=100$ and $t_{\mathrm{\max}}=1000$.}  
  \label{fig:roi_sin}
\end{figure}

Figure~\ref{fig:roi_sin} shows an example of the RoI for different noise level $\sigma$.

%%%%%%%%%%%%%%%%%%%%%%%%%%%%%%%%%%%%%%%%%%%%%%%%%%%%%%

\subsection{Reference Strategies}
\label{sec:ReferenceStrategies}

For comparison purposes, we present in this section different strategies which are used as a reference for the performance of the adaptive strategy.
Note that we could have considered other type of strategies which may lead to a more complete study. 
However, our main goal is to show the performance of the adaptive strategy GACE comparing it with respect to the performance of other strategies for the same investment scenario.
The reference strategies that we selected may be less complex than the adaptive strategy, however, they may have acces to more information about the scenario.

\subsubsection{Strategies with Zero/Complete knowledge}
\label{sec:StrategiesZeroComplete}

For comparison reasons, we present in this section two strategies which represent two simple behaviors for an agent; the first one, called \emph{Constant-Investment-Proportion} (CP), assumes a simple constant minimal investment proportion, whereas the second one, called \emph{Square-Wave} (SW), increases/decreases the investment proportion accordingly to the periodicity of the returns.
In our approach, the CP strategy represents the agent with zero knowledge and zero-intelligence, whereas the strategy SW represents the agent with complete knowledge of the environment.

\textbf{Constant Investment Proportion}

The simplest strategy for an agent would be to take a constant investment proportion for every time step, for simplicity we call this strategy CP:
\begin{equation}
  \label{eq:q0} 
  q(t)=q_{\mathrm{\min}}=\mathrm{const.}
\end{equation}
Since  the  value  of  $q(t)$  is  always fixed,  this  is  not  really  a ``strategy'', but a fixed attitude toward risk and it  plays a role in physics inspired investment models. 
For this model, it has been shown that if a budget-barrier or incomes are assumed, the budget of the agent reaches a stationary distribution in the course of time and the tail of the distribution can be described with a power law function; see \citep{Levy-Solomon96,Sornette-Cont97}.

\textbf{Square Wave Strategy}

The second strategy we consider as a reference is the strategy called \emph{Square-Wave} (SW).
An agent using this strategy invests $q_{\mathrm{\max}}$ during the positive cycle of the periodic return, i.e. where the return has a larger probability to be positive than negative, and invests $q_{\mathrm{\min}}$ otherwise.

It is important to notice that this \textbf{reference strategy} assumes that the agent \textbf{knows} in advance the \textbf{periodicity}, $T$, of the returns. 

For the sake of completeness, we describe this strategy as follows:
\begin{equation}
  \label{eq:qSW}
  q(t) =
  \begin{cases}
    q_{\mathrm{\max}} & t \bmod T < T/2 \\ 
    q_{\mathrm{\min}} & \mathrm{otherwise.}
  \end{cases}
\end{equation}

Other strategies with a similar behavior to this previous may be proposed. 
For example, the strategy to increase the investment proportion only for the time steps where returns are certain to be positive and not for the whole positive period of the returns.
More in detail, this would mean that the agent is considering the worst scenario, which analytically can be expressed as follows:
\begin{equation}
  \label{eq:worstRoI}
  r_{w}(t)=(1-\sigma) \sin \left( \frac{2\pi}{T(t)}\,t \right) - \sigma.
\end{equation}
It can be shown that by solving $r_{w}(t)=0$ for $t$, the range of time steps in a cycle for which the returns are certain to be positive is determined by:
\begin{equation}
  \label{eq:rangeCertain}
  \left[\epsilon, \left( T/2\right)-\epsilon \right],
\end{equation}
where:
\begin{equation}
  \label{eq:rangeEpsilon}
  \epsilon= -\frac{T}{2\pi} \arcsin \left( \frac{\sigma}{\sigma -1} \right).
\end{equation}

\begin{figure}[th]
  \centerline{
    \psfig{file=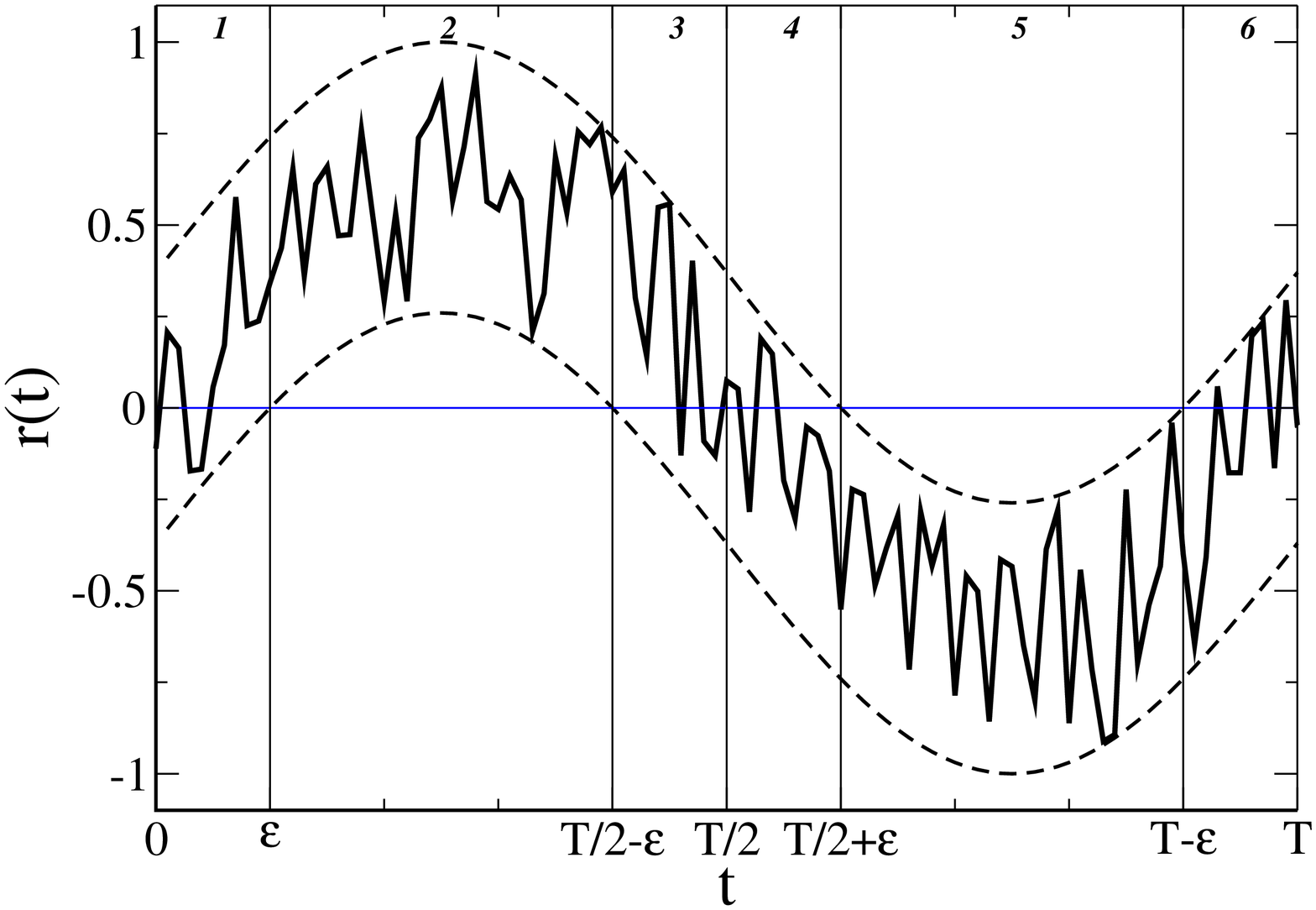,width=8.5cm}
    \psfig{file=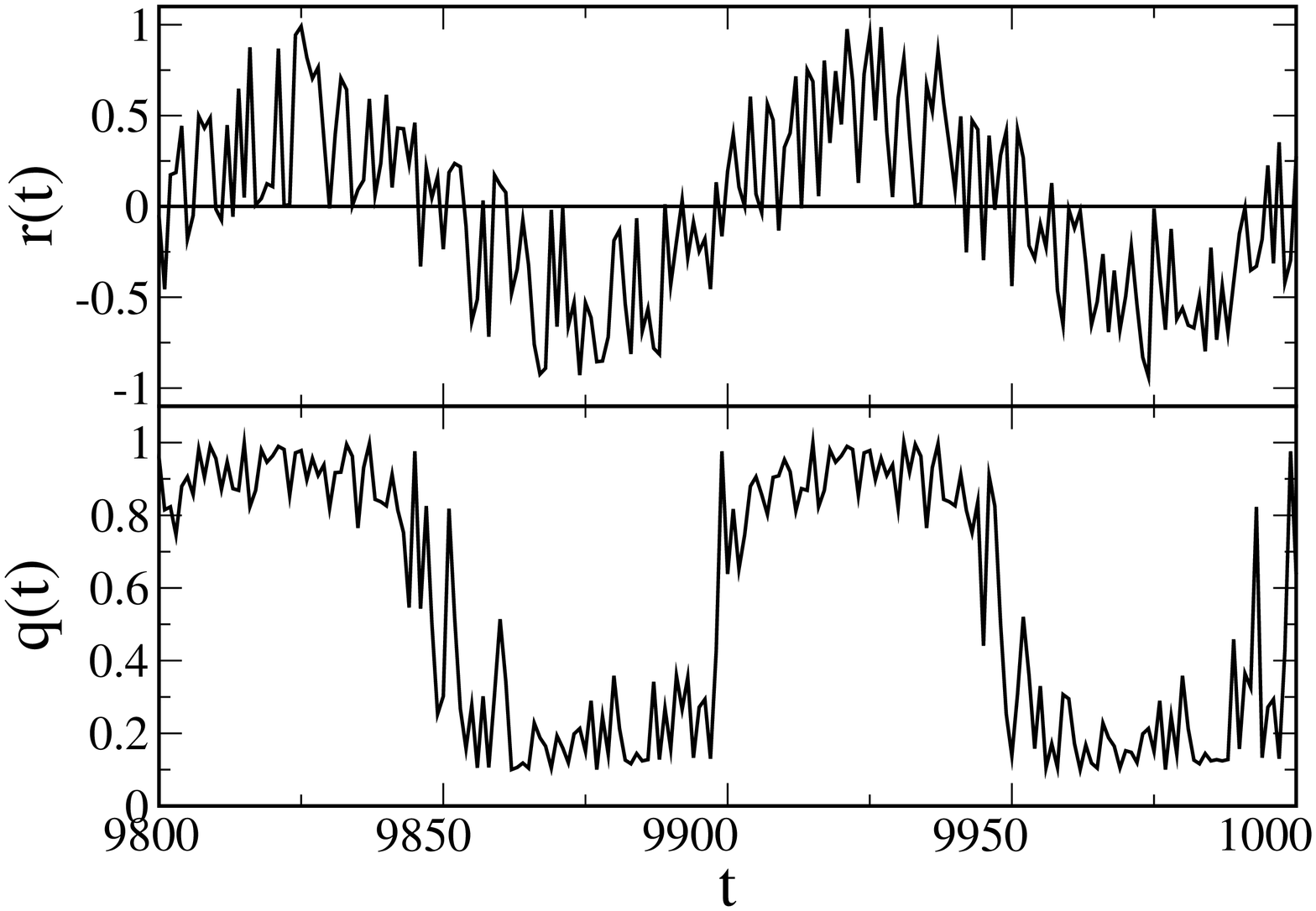,,width=8.5cm}}  
  \vspace*{8pt}
  \caption{Return and investment strategy: (left) Different regions for the returns based on the noise level. Large investments are recommended in region 2 because returns are certain to be positive, and (right) return, $r(t)$ and the investment proportion, $q(t)$, using the strategy GACE after $t=10^4$ time steps. Both for returns with $T=100$ and $\sigma=0.5$}
  \label{fig:RoIAndGA}
\end{figure}

We illustrate this in Figure~\ref{fig:RoIAndGA} (left). 
Note that large investments should be performed only in region 2 where returns are certain to be positive, whereas in regions 1 and 3 the agent may perform only moderate investments, and in the other regions the agent should in general avoid any investment.

In order to elucidate the performance of these previous possibilities for the SW strategy, lets consider returns with no noise, $r(t)=\sin(2 \pi t / T )$, and the agent's wealth dynamic in Eq.~(\ref{eq:wealth}) with initial budget $x(0)=10$. 
For an agent using the investment strategy SW in Eq.~(\ref{eq:qSW}) with $q_{\mathrm{\min}}=0.1$ and $q_{\mathrm{\max}}=1.0$, we find that the maximum possible budget for $T=100$ after $t=100$ time steps is $x(100)=6.746 \times 10^{9}$.
Now, assuming that returns have some noise, $\sigma$, for $T=100$ we find that for $\sigma=0.1$ and after $t=100$ time steps, the strategy SW leads to the budget $x(100)=1.489 \times 10^{9}$.
Note that following Eq.~(\ref{eq:rangeCertain}) for $T=100$ and $\sigma=0.1$, returns are certain to be positive for time steps in the range $t \bmod T \in [2,48]$. 
Now, if we assume an agent with the investment strategy to be a constant investment proportion of $q=1.0$ for time steps in this range and $q=0.1$ otherwise, it can be shown that at the end of a cycle this leads to a budget of $x(100)=1.226 \times 10^{9}$, which is less than the budget obtained using SW.
Furthermore, for $\sigma=0.5$, if (despite the noise) the agent uses again the SW strategy, this leads after $t=100$ time steps to the budget $x(100)=1.373 \times 10^{6}$, whereas if the agent has an investment proportion of $q=1.0$, for those time steps in the ranges: $t \bmod T =\{[2,48], [10,40], [24,26]\}$ and $q=0.1$ otherwise. 
It can be shown that these previous may lead to the budgets $x(100)=\{ 1.23 \times 10^{6}, 116015, 13.41\}$, respectively. 
This means that even returns have large noise, the best strategy is to increase the investment proportion once the returns are more probable to be positive than negative and not only for the returns that are certain to be positive.

\subsubsection{Strategy based on Technical Analysis}
\label{subsec:StrategyBasedOnTechnicalAnalysis}

We decided to include in our study a strategy based on \emph{technical analysis} methods, which are frequently used by traders to forecast returns.

For simplicity, we chose the \emph{Moving Least Squares} (MLS) technique and we avoided strategies based on \emph{Moving Averages} (MA). 
When the latter is used, there is a 'lag' in time with respect to the current return. This causes an underestimation/overestimation for increasing/decreasing returns. 

For the strategy MLS, we consider an agent with a memory size $M$ to store previous received returns, and basically this strategy fits a function to the previous $M$ returns, to estimate the next return, $\hat{r}(t)$. 
For simplicity, we chose this function to be a linear trend-line, which is found by minimizing the distance of this function to the stored returns. 

Noteworthy, once the next return has been estimated, the agent still needs to perform the corresponding adjustment of the investment proportion.
For this, we consider that the agent has a \emph{risk-neutral} behavior, i.e. for small or large fluctuations of the RoI, the agent updates its investment proportion according to the expected return only.
In this approach, the value of $q(t)$ is updated as follows:
  \begin{equation}
  \label{eq:riskNeutralMLS}
    q(t) =
    \begin{cases}
      q_{\mathrm{\min}} & \textrm{$\hat{r}(t) \le q_{\mathrm{\min}}$} \\
      \hat{r}(t) & q_{\mathrm{\min}} < \hat{r}(t) < q_{\mathrm{\max}} \\
      q_{\mathrm{\max}} & \textrm{$\hat{r}(t) \ge q_{\mathrm{\max}}$} 
    \end{cases}
  \end{equation}
  where $q_{\mathrm{\min}}, q_{\mathrm{\max}} \in [0,1]$.   
  In other words, the agent invest $q_{\mathrm{\min}}$ if the estimated return for the next time step is negative or zero, otherwise it invests proportional to the estimated next return.

\subsection{Results for RoI with fixed periodicity}
\label{subsec:RoIWithFixedPeriodicity}

To elucidate the performance of the adaptive strategy proposed in this paper and the reference strategies previously presented, we start with a simple scenario where returns have a fixed periodicity.
In Section \ref{subsec:RoIWithChangingPeriodicity} we consider a more challenging scenario where returns have a changing periodicity. 

First, we assume that the parameters of a strategy lead to an optimal performance, if it leads to the \emph{maximum total budget} that can be reached with this strategy during a complete period of the returns. 
When evaluating the strategies, we have to consider that their performance is also influenced by stochastic effects and. 
In the case of the strategy GACE we also have to account for the different possible strategies that may evolve.
This means that we have to average the simulation over a large number of trials, $N$, where each trial simulates an agent acting independently with the same set of strategy parameters.
More in detail, the performance of an agent in a single trial corresponds to the average budget at the end of each RoI's period, $T$; thereafter, an average over a number of trials is performed to diminish noise effects. 
For convenience, the total budget has been normalized by the number of cycles or periods of the RoI, $I$. 
This is done, because if the strategy performs well, the budget of the agent may reach very high values. 
This occurs because in the dynamics of eqn. (\ref{eq:wealth}) the budget could possibly be doubled at each time step, if an appropriate $q(t)$ and $r(t)$ are provided. 
In the computer simulations this would lead to numerical overflows, therefore we have chosen to reinitialize the budget after each cycle of the RoI, which applies to all simulations, to ensure comparison.

\subsubsection{GACE Parameter Tuning}

In the following, we want to find the parameter values that lead to larger fitness and budget values for the strategy GACE.
It  is well  known that  the  configuration of  most metaheuristic  algorithms requires  both complex  experimental designs  and high  computational efforts.
For finding the best parameters for  the GA, a software called \emph{+CARPS (Multiagent  System for Configuring Algorithms in Real Problem  Solving)} \cite{monett04a}  was used.  It  consists of autonomous,  distributed, cooperative agents  that search  for solutions  to a configuration problem, thereby fine-tuning the metaheuristic's parameters.

The \emph{GA} was configured for periodic returns with $T=100$ and different level of noise: $\sigma=0.1$,  and $\sigma=0.5$. 
In  this  process,  four  GA parameters  were  optimized:  the population size $C$, the crossover  probability $p_{c}$, the mutation probability $p_{m}$, and the elitism size $s$. 
Their  intervals of definition, in which  the most acceptable GA configurations should be found, were set as follows: $C \in \{ 50, 100, 200,  500,  1000\}$,  $p_{c}  \in  \left[0.0,  1.0\right]$,  $p_{m}  \in \left[0.0,   1.0\right]$,   and   $s   \in   \left[0.0,   0.5\right]$. 

When  configuring,  agents in  +CARPS  apply  a  Random Restart  Hill-Climbing approach    and   they    exchange   best-so-far    solutions    during   this process. 
Furthermore, the evaluation of the GA with a particular configuration is repeated  five times in  order to cope  with its stochastic  nature.  
According to the fitness, we show in Table~1 the best obtained configuration for the GA in the periodic returns previously mentioned.
\begin{table}[h]
  \centering 
  \caption{GACE's best parameter values for RoI with fixed $T$.} 
  \label{table:soluts} 
  \begin{tabular}{|c|c|c|c|} 
    \hline 
    $C$ & $p_{c}$ & $p_{m}$ & $s$ \\ 
    \hline 
    1000 & 0.7 & 0.01 & 0.3 \\ 
    \hline
  \end{tabular}
\end{table}

\begin{figure}[th]
  \centerline{
    \psfig{file=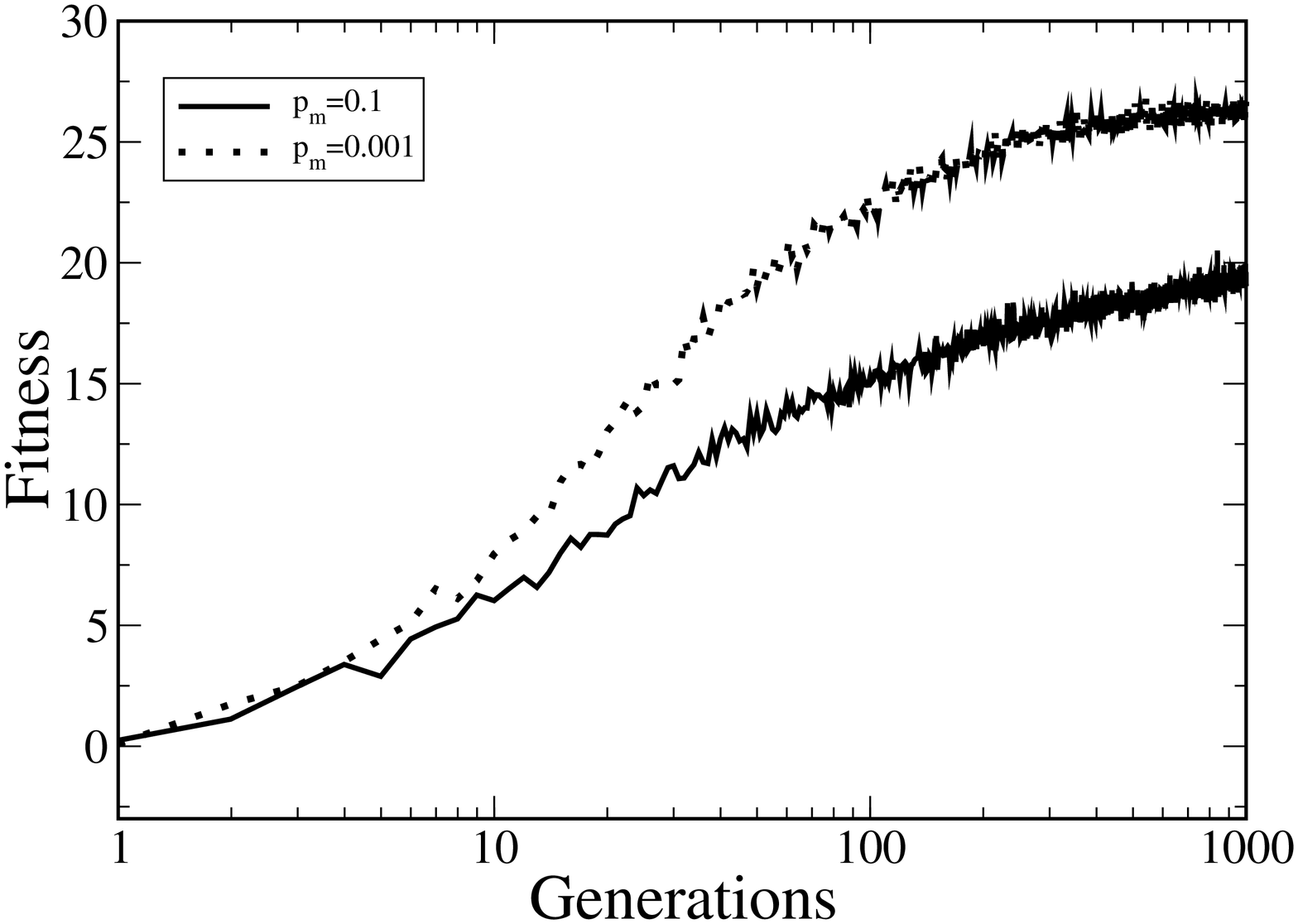,width=8.5cm}
    \psfig{file=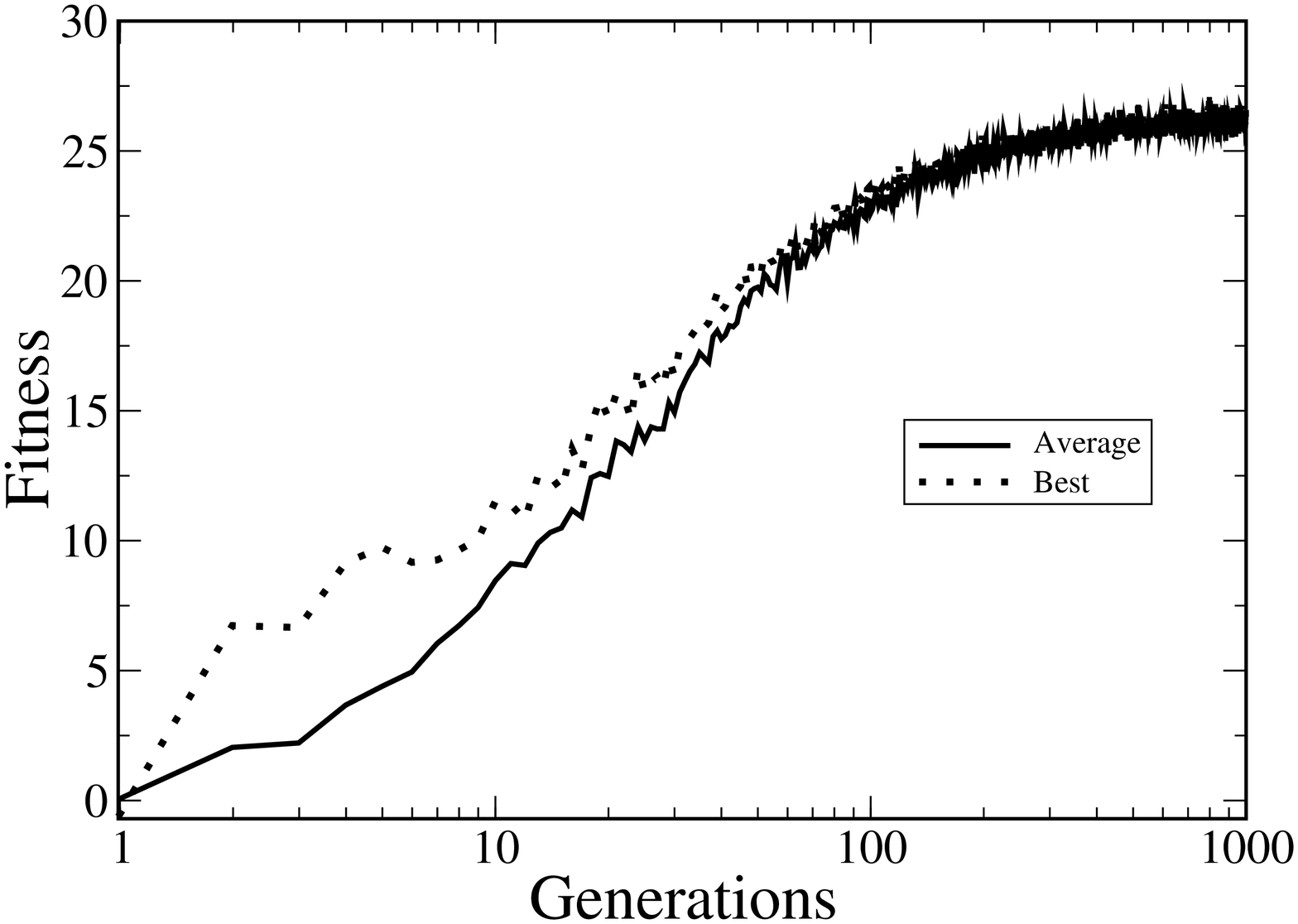,width=8.5cm}}
  \vspace*{8pt}
  \caption{Performance graphs for the GACE strategy in the course of generations for $C=1000$ chromosomes: (left) average fitness for different mutation probabilities, $p_m$; and (right) convergence of the population showing the average fitness against the fitness of the best chromosome for $p_m=0.01$. Further parameters $T=100$ and $\sigma=0.1$.}
  \label{fig:FitnessvsTauSDAMP}
\end{figure}

For the sake of completeness, we show in Figure~\ref{fig:FitnessvsTauSDAMP} (left)  the evolution in the course of generations of the average fitness of the chromosomes in the population for different mutation rates.

Moreover, in Figure~\ref{fig:FitnessvsTauSDAMP} (right) shows for $C=1000$ chromosomes, the evolution of the average fitness and the largest fitness in the course of generations.
Observe that the rate $p_m=0.01$ used for Figure~\ref{fig:FitnessvsTauSDAMP} (right), leads to larger average fitness in the population than for less or more mutation rate showed in Figure~\ref{fig:FitnessvsTauSDAMP} (left); however, note that the fitness of the best chromosome when using $p_m=0.001$ is almost as well as for $p_m=0.01$.

Note that in Figure~\ref{fig:FitnessvsTauSDAMP} (right), for the first 100 generations the best chromosome performs much better than all the chromosomes in average; however, after 100 generations we can see that the performance of the population converges to the performance of the best chromosome.
Now, consider again Eq.~(\ref{eq:fitness}) and replace: $g_{jk}$ with $q(t)$, and $G_{j}$ with $T$.
If we consider returns for $t=100$ time steps with periodicity $T=100$ and no noise, it can be shown that the strategy SW would lead to a fitness of $f(\tau)=28.63$.
Note that this is not much larger than the fitness obtained with GACE.

Now, to better illustrate the set of investment strategies that are being obtained using GACE, we show in Figure \ref{fig:RoIAndGA} (right) the RoI and the investment proportions obtained after a number of time steps for returns with relative large noise.
For the reader with background in signal processing techniques, Figure \ref{fig:RoIAndGA} may sound familiar as it resembles to those figures obtained when using matched filters for signal recovery, see \citep{turing60} for more on matched filters.

\subsubsection{Performance Comparison}
\label{subsubsec:PerformanceComparison}

\begin{figure}[th]
  \centerline{
    \psfig{file=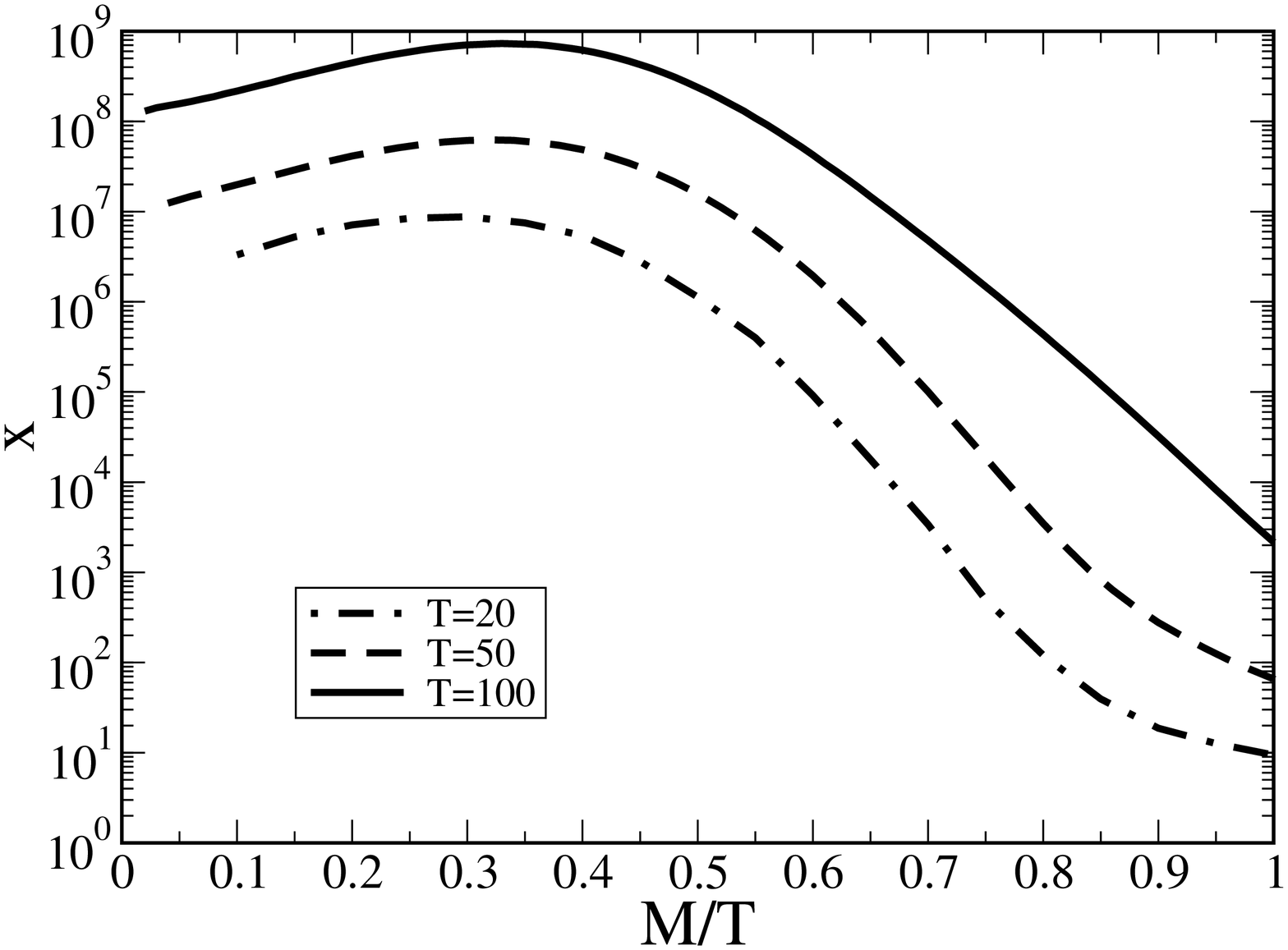,width=8.5cm}
    \psfig{file=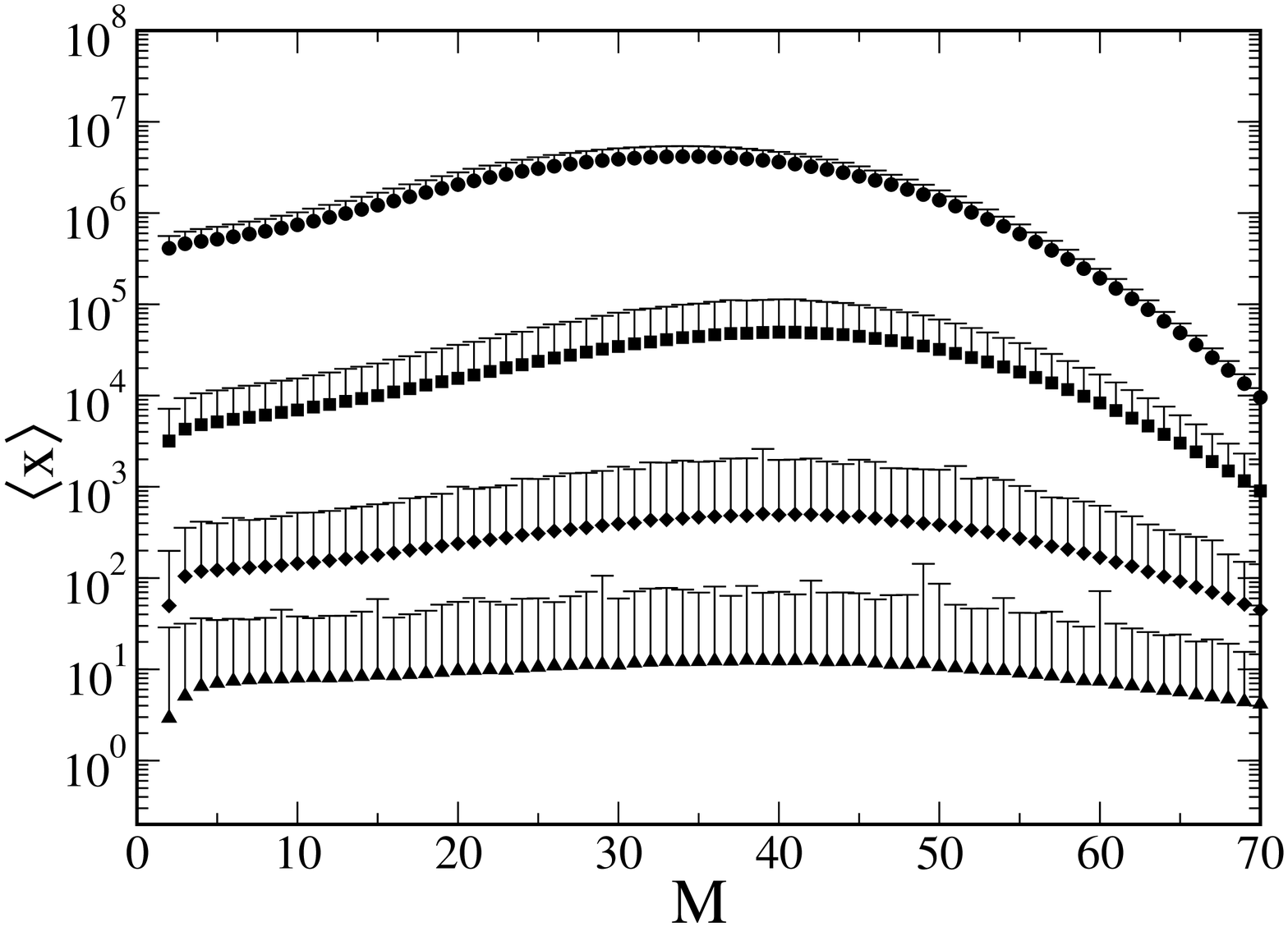,width=8.5cm}}
  \vspace*{8pt}
  \caption{For strategy \emph{MLS}, both plots show the budget obtained for different memory size $M$ for: (left) different periodicity $T$ and no noise, (right) different noise level $\sigma$ and periodicity $T=100$.}
  \label{fig:MLSXvsMdivTdiffT} 
\end{figure}

In order to assure fair comparison between the strategies, we need to find the most proper parameter values for the strategies.
Note that for both strategies CP, Eq.~(\ref{eq:q0}) and SW, Eq.~(\ref{eq:qSW}), we don't need to tune any parameters.
However, for the strategy MLS, Eq.~(\ref{eq:riskNeutralMLS}), we assume that the agent has acces to some information about the returns, i.e. the agent knows the periodicity, $T$, of the returns.
This means that the agent needs to determine the most proper memory size, $M$, based on the known periodicity of the returns.
For this, we performed some experiments for returns with different fixed periodicities, $T$, and no noise, $\sigma=0$.
Figure~\ref{fig:MLSXvsMdivTdiffT} (left), shows the results of these experiments, where the budget of an agent is shown for different memory sizes and for returns with different $T$, and no noise.
According to visual impression, the most proper memory size, $M$, and the periodicity, $T$, are proportionally related by $M/T \approx 0.37$.

Moreover, if we assume returns with no noise, we can find analytically the memory size $M^{\star}$ that maximizes the profits.
For this, we note that for a periodic return as in Eq.~(\ref{eq:roi_sin}) with $\sigma=0$, the strategy MLS estimates the next return $\hat{r}(t+1)$ as follows:
\begin{eqnarray}
  \label{eq:mlsestimate}
  \hat{r}(t+1)&=&\frac{\sin \left(\omega\,t \right)-\sin \left( \omega\,(t-M) \right)}{\omega\, t -\omega \left( t-M\right)} \left( \omega(t+1)-\omega(t-M)\right)+\sin \left(\omega t- \omega M\right) \nonumber \\ 
  &=&\frac{M+1}{M} \left[ \sin \left(\omega\,t \right)- \sin \left( \omega t- \omega M\right)\right],
\end{eqnarray}
where $\omega=2\pi/T$.
Now, by calculating the average profits $\mean{r\, q}$ for the positive cycle of the returns, we find:
\begin{eqnarray}
  \label{eq:mlsprofit}
  \mean{r\, q} &=& \int_{0}^{T/2} r(t)\, q(t) dt \nonumber \\
  &=& \frac{T \left(M+1-\cos \left(\omega\, M \right)\right)}{4\,M}.
\end{eqnarray}

Figure~\ref{fig:MLSXvsMdivTdiffT} (right) shows the resulting budgets for different memory size values when using Eq.~(\ref{eq:mlsprofit}). 
Note that we can find the memory size that leads to maximum profits by finding the derivative of $\mean{r\, q}$ w.r.t $M$, which is:
\begin{equation}
  \label{eq:mlsprofitpartial}
  \partial_{M}\mean{r(t)\,q(t)}=  \frac{-T \sin\left( \frac{\omega}{2}\,M\right)^2+\pi\,M\, \sin \left(\omega \,M\right)}{2\,M^2}.
\end{equation}

Thus, the memory size $M^{\star}$ that maximizes the profits can be calculated by solving $\partial_{M}\mean{r(t)\, q(t)}=0$. 
Using Taylor series to the sixth order for the sinusoidal functions we end with the following expression:
\begin{equation}
  \label{eq:optmlsM}
  M^{\star}=\frac{\sqrt{\frac{3}{2}}}{\pi} \, T.
\end{equation}
Consequently, for $T=100$ the theoretical optimal memory size is $M^{\star}=38$, which agrees with the empirical optimal memory size shown in Figure \ref{fig:MLSXvsMdivTdiffT} (right) for different noise levels.
Relatedly, the proportion $M/T \approx 0.37$ found by means of computer simulations in Figure~\ref{fig:MLSXvsMdivTdiffT} (left), approximates pretty well the proportion found analytically $M/T = \sqrt{\frac{3}{2}}/\pi = 0.389$.

\vspace{0.3cm}

\textbf{Comparison for fixed chromosome length}

\vspace{0.3cm}

Now, we compare the performance of the adaptive investment strategy GACE, presented in Section~\ref{subsec:StrategyBasedOnEvolution}, with respect to the reference strategies presented in Section~\ref{sec:ReferenceStrategies}.
For the sake of clarity, we assume for the moment that the strategy GACE uses fixed chromosome length, i.e. $G_{j}=G_{\mathrm{\max}}$.
For all strategies we consider $q_{\mathrm{\min}}=0.1$ and $q_{\mathrm{\max}}=1.0$ in our experiments.
These parameter values describe the behaviour of the strategies CP, Eq.~(\ref{eq:q0}), and SW, Eq.~(\ref{eq:qSW}).
For the strategy MLS, Eq.~(\ref{eq:riskNeutralMLS}), we use Eq.~(\ref{eq:optmlsM}) to determine the optimal memory size and for the strategy GACE we use the parameters in Table~1.

As it was done previously, we generate a synthetic data set for the returns. 
In our experiments we assume that the agent invests in returns with periodicity $T=100$ for different noise levels.
We consider here that the length of the chromosomes is fixed to $G_j=100$ and a new generation of chromosomes is being obtained after a number of time steps $t_{\mathrm{eval}}=100$.
In other words, we consider for these experiments the approach GMaximum to determine the number of time steps needed to evaluate the population, see Section~\ref{sec:TimeForEvaluation}.
Note that we investigate the simplest case where the periodicity of the returns perfectly maps to the length of the chromosomes and time steps for evaluation of the population.
For the computer experiments, we let the agent to use one of the strategies to invest during a number of $t=10^5$ time steps.
In order to account for the randomness of the scenario, we perform the experiment for a number of $N=100$ trials, gathering the average budget obtained for each strategy at every 100 time steps.

\begin{figure}[th]
  \centerline{
    \psfig{file=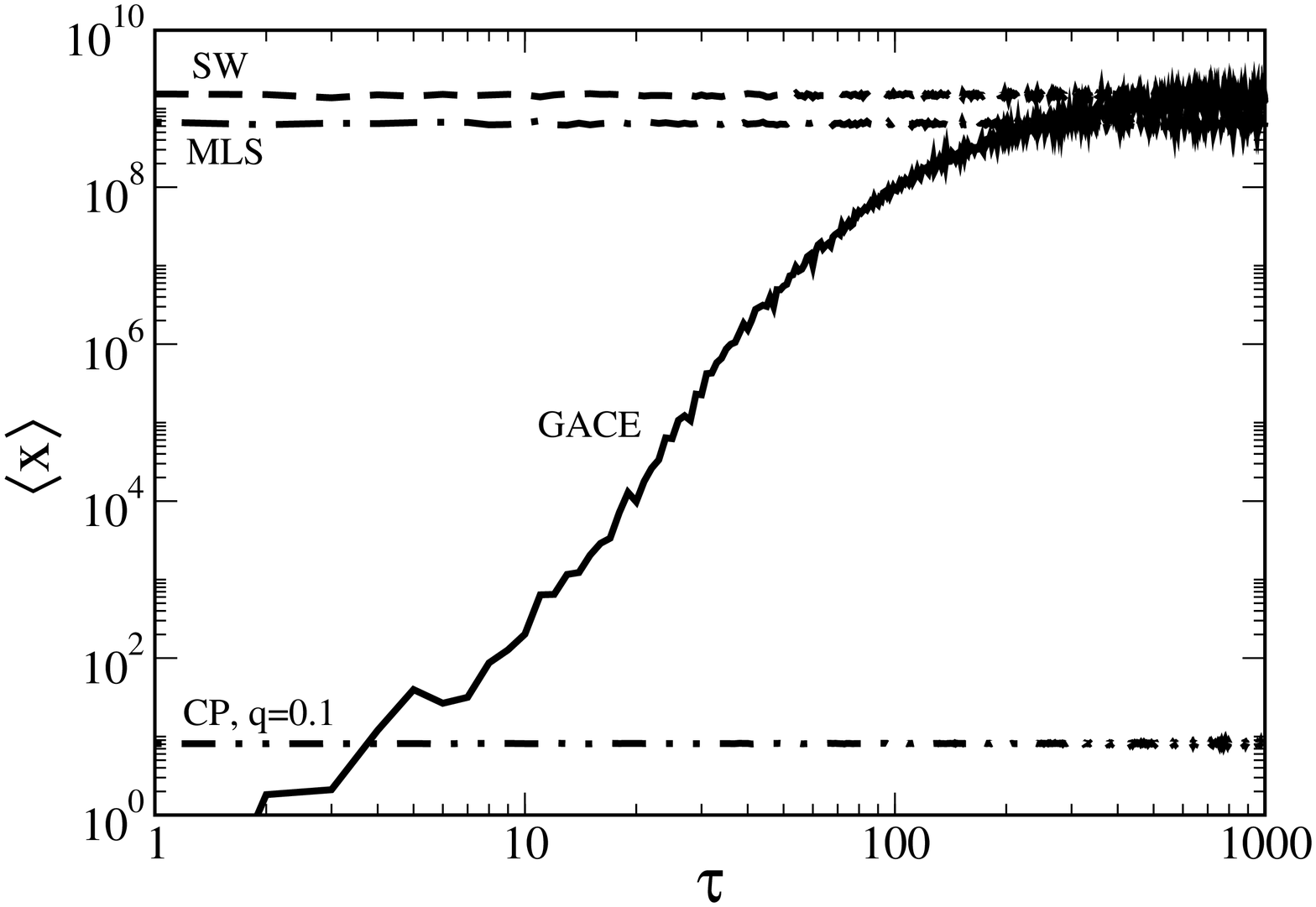,width=8.5cm}
    \psfig{file=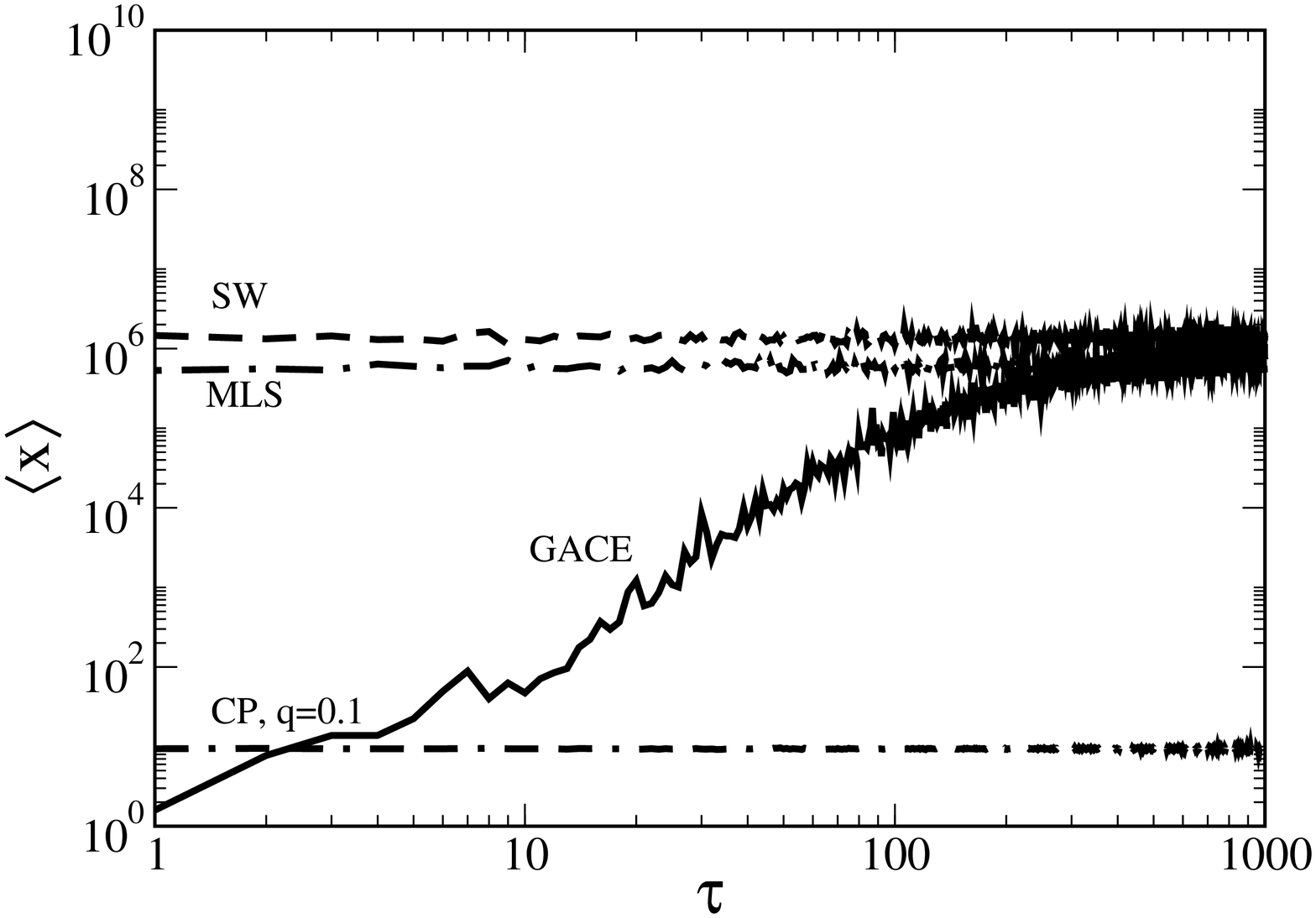,width=8.5cm}}
  \vspace*{8pt}  
  \caption{Average budget, $\mean{x}$, obtained using different investment strategies in the course of  generations $\tau$ of the strategy GACE, for returns with periodicity $T=100$ and amplitude noise: (left) $\sigma=0.1$ and (right) $\sigma=0.5$.}
  \label{fig:BudgetGACEMLSQ0_1RRvsTau}
\end{figure}

Figure~\ref{fig:BudgetGACEMLSQ0_1RRvsTau} shows in a log-log plot the average budget, $\mean{x}$, in the course of GACE's generations, $\tau$, for all strategies and for returns with different amplitude noise levels.
As you can see, except for the GACE strategy, all other strategies have a constant budget in average over each generation. 
This occurs because the average of the budget was taken at every $t_s=100$ time step which corresponds to the periodicity of the returns $T=100$ and to the time steps to evaluate the population of chromosomes $t_{\mathrm{eval}}=100$, as it was specified in our experiment parameters.

More in detail, Figure~\ref{fig:BudgetGACEMLSQ0_1RRvsTau} (left) shows that after 4, 70, and 300 generations, GACE over-performs the strategies $q=0.1$, and MLS, respectively.
And we note that GACE performs almost as well as the strategy SW after 400 generations.
Moreover, the budget of the agent using GACE increases approximately according to a power law for the first 100 generations, afterwards increases logarithmically.

Figure~\ref{fig:BudgetGACEMLSQ0_1RRvsTau} (right) shows that for large amplitude noise it takes fewer generations for GACE to over-perform the strategy $q=0.1$, but in general more generations are needed for GACE to over-perform the other strategies. 
We find that the budget also increases according to a power law for the first 100 generations and afterwards increases logarithmically.
We note that it would be useful to provide with a formulation to characterize the average budget in the course of generations that is obtained using the strategy GACE, however, this is left for further work.

\vspace{0.3cm}

\textbf{Comparison for variable chromosome length}

\vspace{0.3cm}

In this section, we investigate the case when the chromosomes do not have initially the same length as the periodicity of the returns.
For this, we let the initial length of the chromosomes to be chosen at random from a Uniform distribution.
This means that we need to define the range of possible length values.
For our implementation of the GA to work properly, the unkown periodicity needs to be in the range of possible length values.
For simplicity, in our experiments we let the range to be larger than the periodicity of the returns.
However, we note that this parameter could be determined by the GA itself if we include extra genes in the chromosome to track for a proper range.
Another possibility could be to determine this parameter by means of statistical properties of the returns, like the autocorrelation function or spectral density; however, both approaches are beyond the scope of this paper and are left for further work.

Now, for the case that initially the length of the chromosomes is different, if population is evaluated after a fixed number of time steps, the following question may arise: \emph{Do the chromosomes' lengths correctly evolve to map the periodicity of the returns?}

To answer this question, we performed some computer experiments for an agent using the strategy GACE for returns with periodicity $T=100$ and different noise levels. 
For these experiments, we assumed for GACE the parameter values specified in Table~1, and of special interest, we now consider that the initial chromosomes' length is drawn randomly from a Uniform distribution with range of values $(1,G_{\mathrm{\max}})$, with $G_{\mathrm{\max}}=500$. 
Furthermore, we assume that the evaluation of the population, leading to a new generation of chromosomes, is performed every $t_{\mathrm{eval}}=500$ time steps, i.e. we consider for these experiments the approach $G_{\mathrm{\max}}$ to determine the number of time steps needed to evaluate the population, see Section~\ref{sec:TimeForEvaluation}.

\begin{figure}[th]
  \centerline{
    \psfig{file=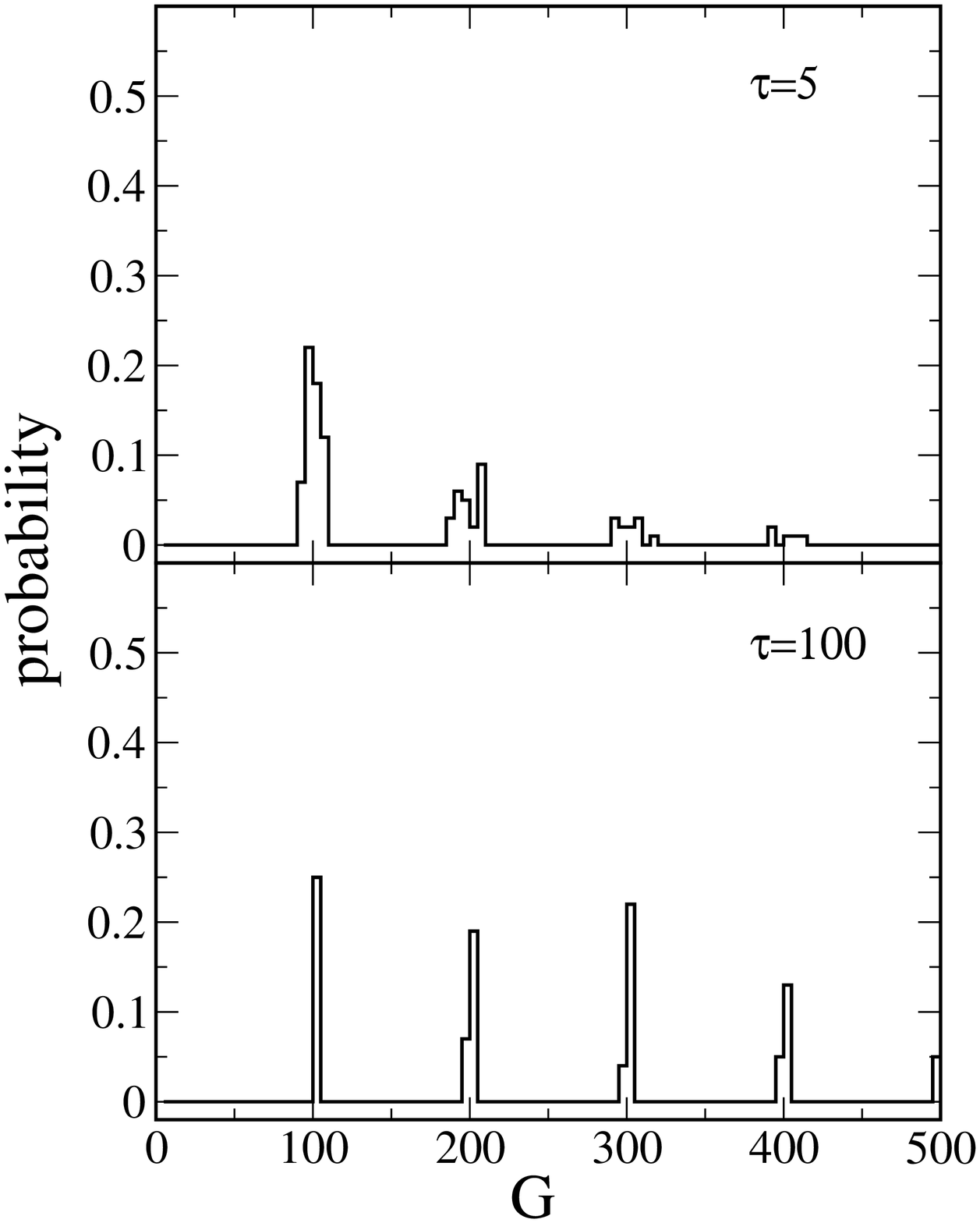,width=8.5cm}
    \psfig{file=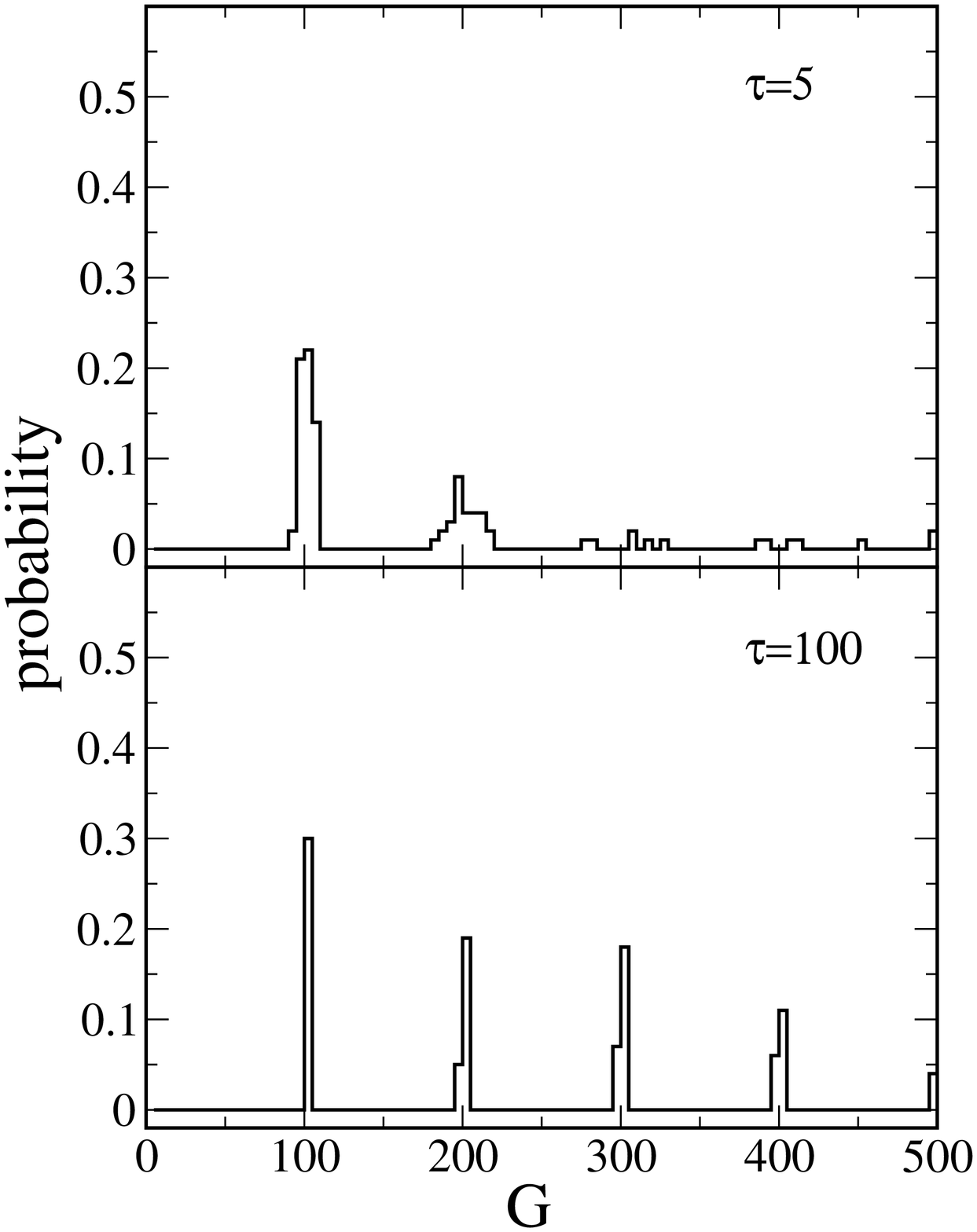,width=8.5cm}}
  \vspace*{8pt}
  \caption{Probability distribution of the length of the best fitted chromosomes for generations $\tau=\{5,100\}$ for $N=50$ trials. Returns with periodicity $T=100$ and amplitude noise: (left) $\sigma=0.1$ and (right) $\sigma=0.5$.}
  \label{fig:distgenesGCrossResc}
\end{figure}

Figure \ref{fig:distgenesGCrossResc} shows the probability distribution of the length of the best fitted chromosomes for different noise levels and for different generations, $\tau=\{5,100\}$.

It is clear that after five generations most of the chromosomes' length have properly matched the periodicity of the returns.
Interestingly, chromosomes with lengths proportional to a multiple of the periodicity are also frequent; however, the probability decreases for larger multiples of the real periodicity, which is a consequence of the better adaptation of smaller chromosomes which have found more quickly the most proper investment proportions.

\subsection{RoI with changing periodicity}
\label{subsec:RoIWithChangingPeriodicity}

In the previous section, we deal with a stationary environment, now in this section we tackle a non-stationary environment.
For comparison reasons, we start presenting some computer experiments for the strategy GACE using the parameters for a stationary environment, shown in Table 1, now for returns with non-fixed periodicity.

\begin{figure}[th]
  \centerline{
    \psfig{file=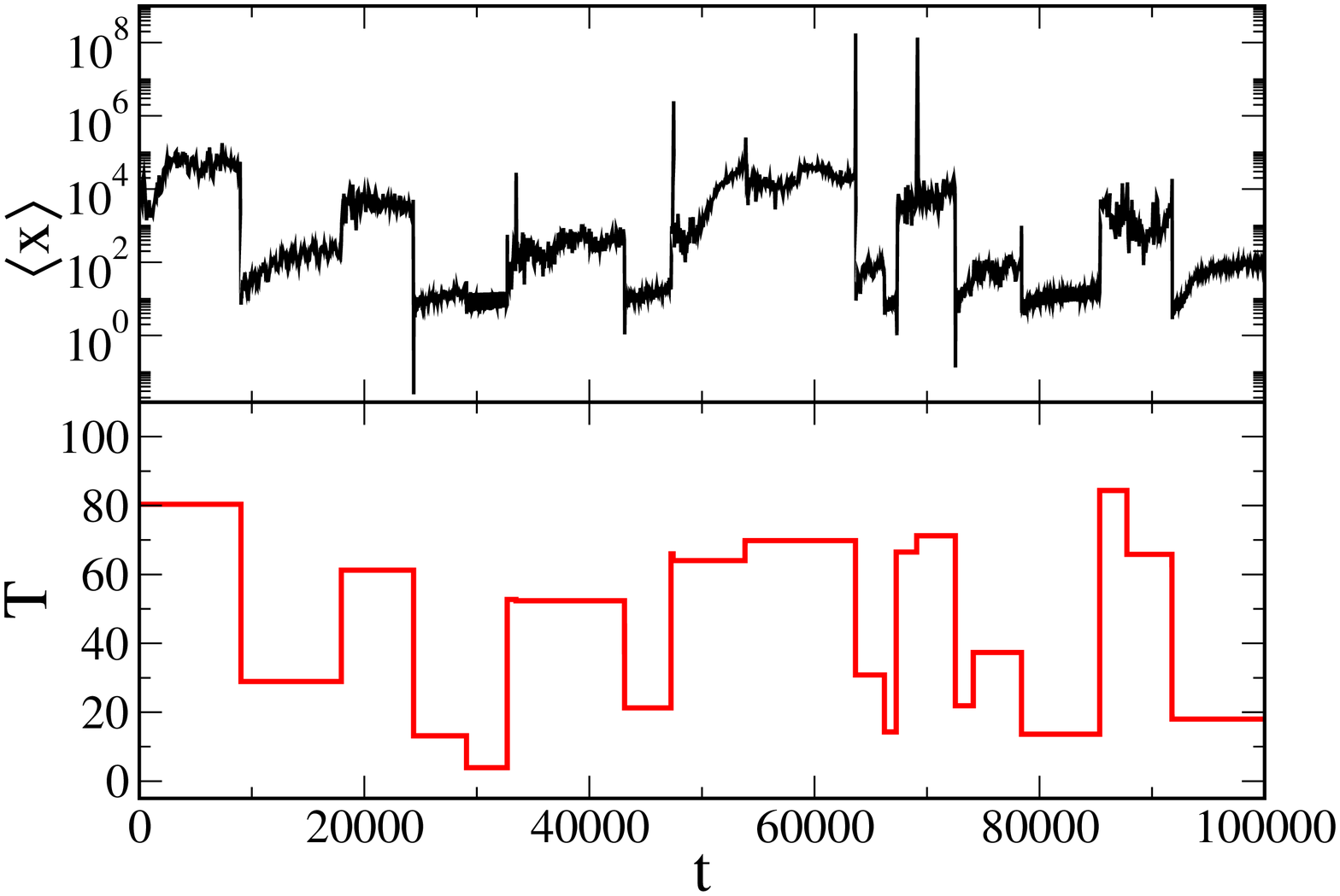,width=8.5cm}
    \psfig{file=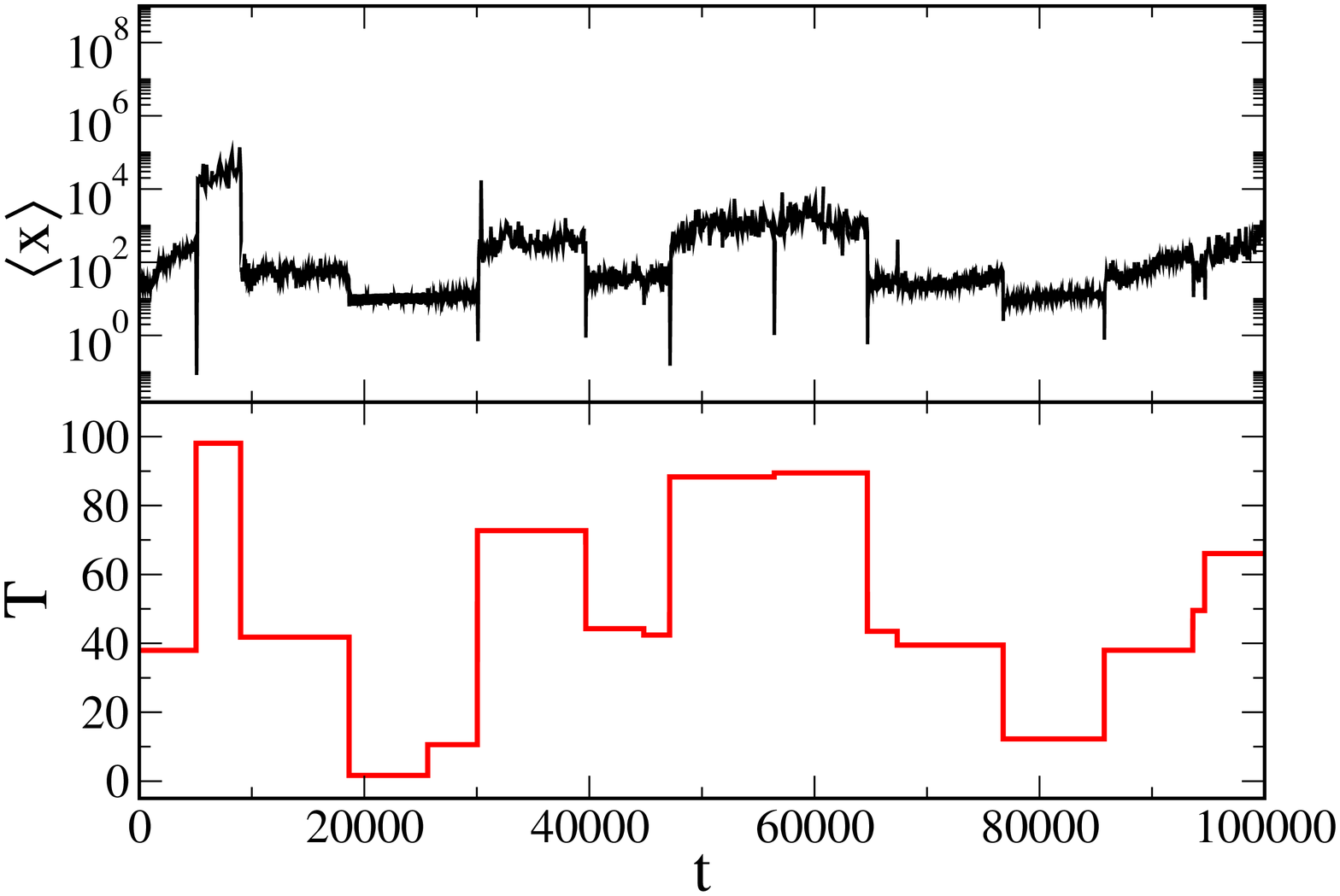,width=8.5cm}}
  \vspace*{8pt}
  \caption{(top) Average budget in the course of time for $N=50$ trials for an agent using the strategy GACE with parameter values as in Table 1 and the parameter values $G_{\mathrm{\max}}=200$ and $p_l=0.1$. (bottom) Periodicity of the returns in the course of time, Eq.~(\ref{eq:roi_sin}), with parameters: $T_{\mathrm{max}}=100$ and $t_{\mathrm{max}}=10^{4}$.
    Both for different amplitude noise: (left) $\sigma=0.1$ and (right) $\sigma=0.5$.
  }
  \label{fig:GACEXvstVarTVarG_xmean_roi_DiffSDAMP}
\end{figure}

Figure \ref{fig:GACEXvstVarTVarG_xmean_roi_DiffSDAMP} (top) shows the evolution of the average budget in the course of time for an agent with the strategy GACE investing in returns with changing periodicity and different noise levels.
For the sake of clarity, we include in Figure \ref{fig:GACEXvstVarTVarG_xmean_roi_DiffSDAMP} (bottom) the corresponding periodicities of the returns for each time step.

Note that for these experiments we use the selection approach GBestCurrent, see Section~\ref{sec:TimeForEvaluation}.
Also note that in order to avoid overflows, see Section~\ref{subsec:RoIWithFixedPeriodicity}, the budget of the agent is reinitialized to the initial budget every time the periodicity of the returns changes.
From the dynamics of the returns, Eq.~(\ref{eq:roi_sin}), it can be seen that a change of periodicity is not performed exactly at the end of a period but at any time step. 
This is the reason for large increases or decreases of budget each time the periodicity of the returns changes.

\subsubsection{GA Parameter Tuning}

As we did before, we address the problem of determining the most proper parameter values for GACE, now for returns with changing periodicity.
For this, we performed some experiments and determined empirically the most proper parameter values for GACE when using the approach GBestCurrent, these results are shown in Table~2.
\begin{table}[h]
  \centering 
  \caption{GACE's best parameter values for RoI with changing $T$.} 
  \label{table:soluts} 
  \begin{tabular}{|c|c|c|c|c|} 
    \hline 
    $C$ & $p_{c}$ & $p_{m}$ & $s$ & $p_{l}$\\ 
    \hline 
    1000 & 0.5 & 0.001 & 0.3 &0.5 \\ 
    \hline
  \end{tabular}
\end{table}

Note that with respect to Table~1, if the parameter values in Table~2 are used, the crossover and mutation operators are less probable to occur when recombining two parents.
However, this is covered by a surprising large probability of mutation on the length of a chromosome.

\subsubsection{Performance Comparison}
\label{subsubsec:PerformanceComparison2}

In this section we investigate the performance of the adaptive strategy with respect to the reference strategies in a non-stationary scenario.
For this, we performed some computer experiments for returns with changing periodicity and different noise level.
As we did in the previous sections we assumed for all strategies the parameter values $q_{\mathrm{min}}=0.1$ and $q_{\mathrm{max}}=1.0$.
Moreover, for the strategy MLS we used Eq.~(\ref{eq:optmlsM}) to calculate the memory size, $M$. 
And for the strategy GACE we used the parameter values listed in Table~2 and the length of a chromosome in the range $G_j\in(1,G_{\mathrm{\max}})$, with $G_{\mathrm{max}}=200$.

\begin{figure}[th]
  \centerline{
    \psfig{file=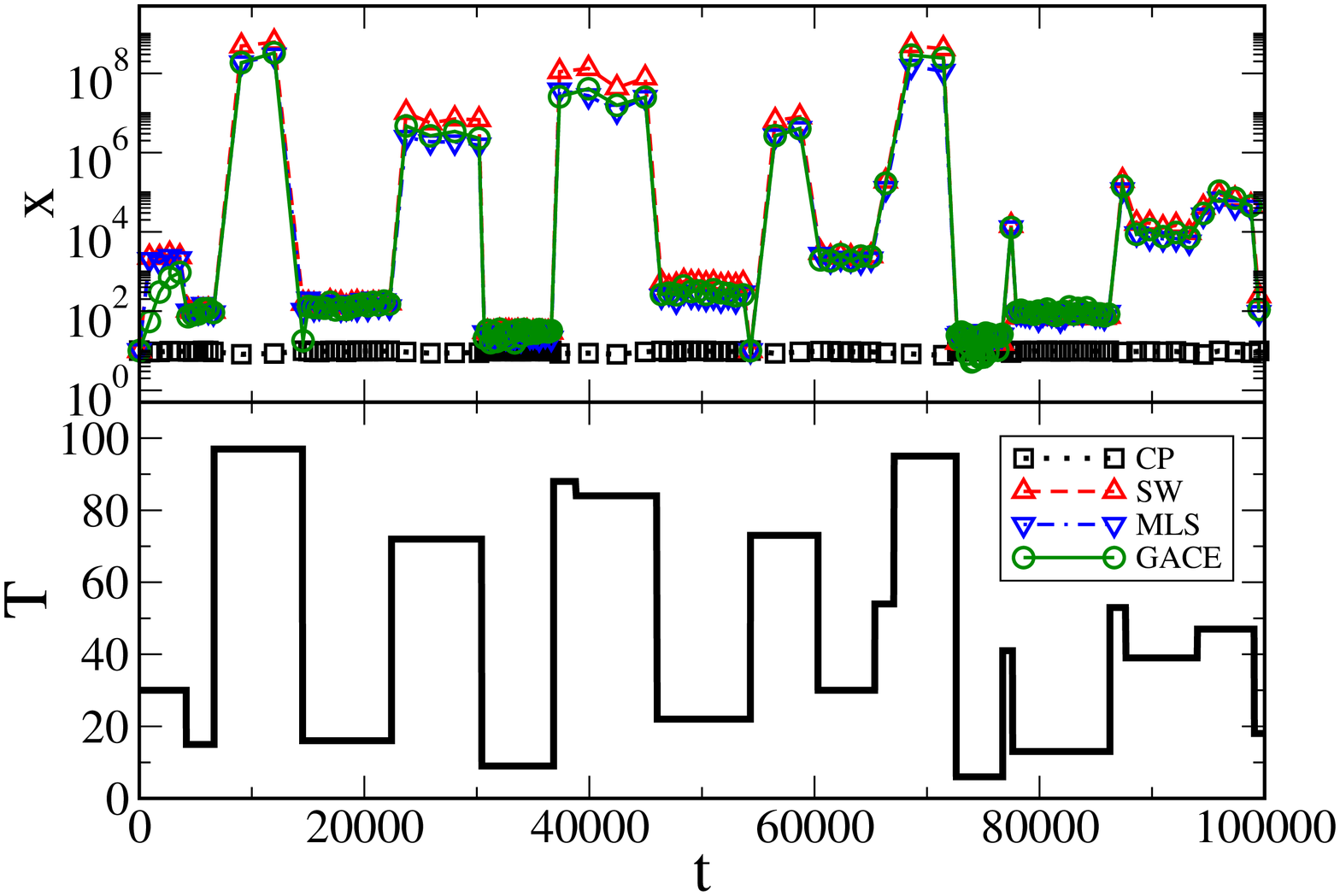,width=8.5cm}
    \psfig{file=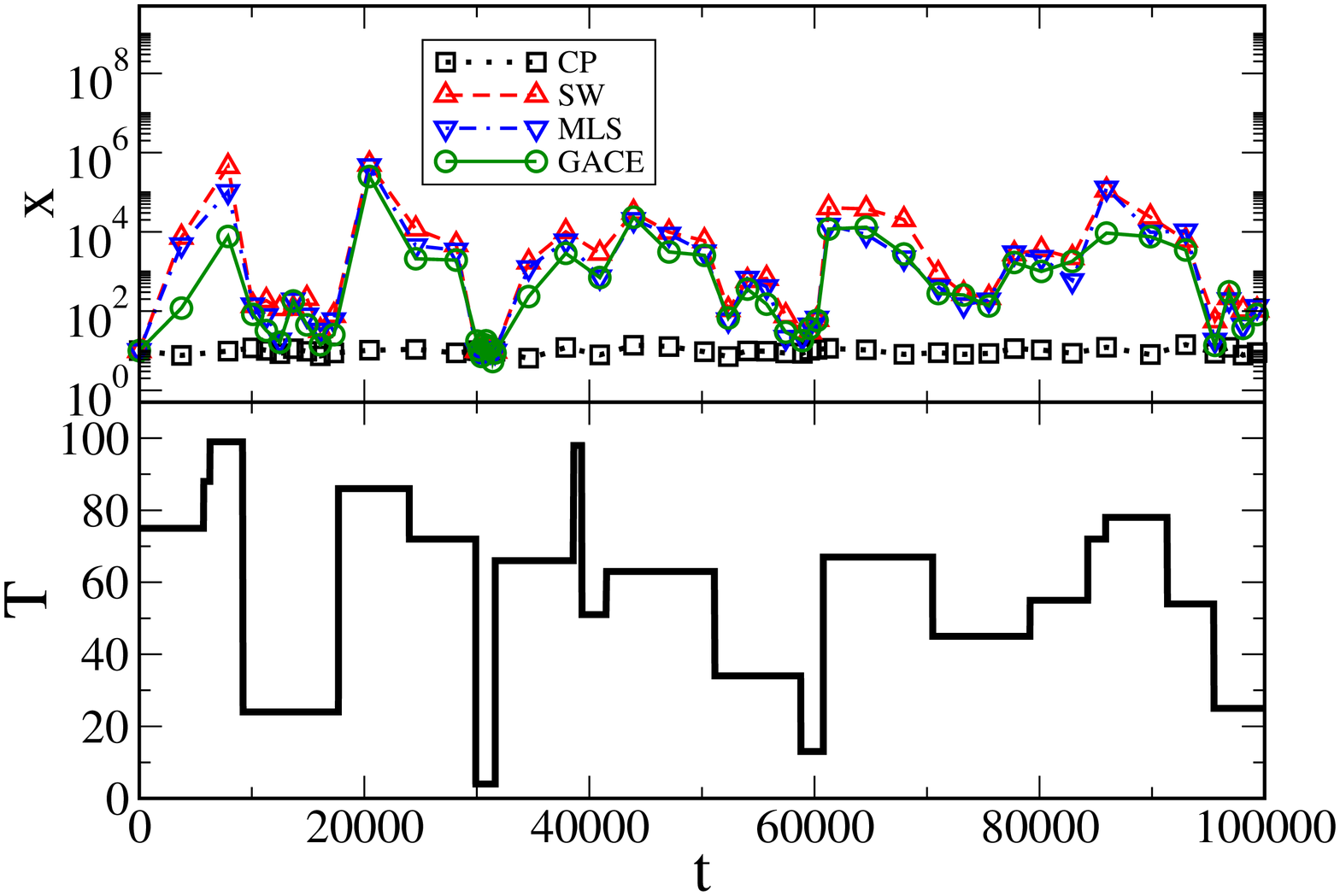,width=8.5cm}}
  \vspace*{8pt}  
  \caption{(top) Budget in the course of time for different strategies, assuming $q_{\mathrm{\min}}=0.1$ and $q_{\mathrm{\max}}=1.0$ for all strategies. For MLS we used Eq.~(\ref{eq:optmlsM}) to calculate the memory size, $M$, and for GACE the parameters shown in Table~2.
    (bottom) Periodicity of the returns in the course of time, Eq.~(\ref{eq:roi_sin}), with parameters: $T_{\mathrm{max}}=100$, $t_{\mathrm{max}}=10^{4}$, and amplitude noise: (left) $\sigma=0.1$ and (right) $\sigma=0.5$.
  }
  \label{fig:AllDiffSDAMP}
\end{figure}

We show in Figure \ref{fig:AllDiffSDAMP} (top) the evolution of budget, and (bottom) the corresponding periodicity of the returns, Eq.(\ref{eq:roi_sin}), both in the course of time for the different investment strategies and different noise levels.
It is clear that the best strategy is for both cases the strategy SW, following the strategy MLS; however, note that both strategies have total and partial knowledge about the dynamics of the returns, respectively.
As we mentioned previously, the strategy SW, Eq.~(\ref{eq:qSW}), knows the dynamics of the stylized returns and increases the investment proportion for the positive periods and decreases it for the negative.
On the other hand, the strategy MLS, Eq.~(\ref{eq:riskNeutralMLS})), knows the periodicity $T$ of the returns, which is used to calculate the most proper memory size by means of Eq.~(\ref{eq:optmlsM}).
This previous knowledge gives some advantage to these strategies over the strategy GACE, which only needs the specification of $G_{\mathrm{\max}}$.
We note that the strategy GACE evolves quite fast, yielding a set of investment strategies with a clear tendency to lead more gains than losses. 
This particularly is shown for long-lasting periodicities, where an ever increasing growth of budget is observed. 
Interestingly, the strategy GACE performs much better than the reference strategy CP and performs on certain occasions as good as the strategy MLS, particularly for returns with small noise.

%%%HERE!!! maybe to add also investment proportion obtained for a different artificial return sin(Ax)+cos((1-A)x)

%%%%%%%%%%%%%%%%%%%%%%%%%%%%%%%%%%%%%%%%%%%%%%%%%%%%%%
\section{Conclusions}
\label{sec:Conclusions}

In this paper, we presented a simple investment model and some investment strategies to control the proportion of investment in periodic environments.
The novelty of this paper is in the adaptive investment strategy here proposed, called \emph{Genetic Algorithm for Changing Environments} (GACE), which is a new approach based on evolution for the correct mapping of investment proportions to patterns that may be present in the returns.
We analyzed the performance of GACE for different scenarios, and compared its performance in the course of time with respect to other strategies that were used as a reference.
We showed that after a given number of time steps, the strategy GACE reaches a set of investment strategies that can over-perform simple strategies like those that invest always a constant investment proportion.
We showed that even though the strategy GACE has no knowledge of the dynamics of the returns, it may lead to large gains, performing as well as other strategies with some knowledge.
This particularly is shown for long-lasting periodicities, where an ever increasing growth of budget was observed. 
This means that in the presence of long-lasting periodicities, the longer the agent uses the adaptive strategy the largest the profits per cycle.

In this study, we used artificial generated stylized returns, which are based on a sinusoidal function; however, it can be shown that for other type of periodic functions, the GA would eventually find the most proper strategy in the same way that for the sinusoidal function.
Despite the fact that the strategy GACE proposed in this paper was mainly used to find the most proper set of investment proportions for an investment scenario, it is important to note that this strategy can be applied to other kind of scenarios.
For example, scenarios where the agent has to control other kind of resources, like energy, time consumption, etc. 

Further work includes the analysis of the performance of the strategy GACE for real returns, and to compare the performance of GACE with other similar approaches like Genetic Programming techniques, Neural Networks, and Reinforcement Learning.
Useful, would be to extend also this approach for optimal portfolio diversification, where a large number of algorithms have been proposed, which deal with the research areas of optimization, stochastic simulation and decision theory.

Finally, we note that the proposed adaptive investment strategy may be interesting for the research area of pattern recognition of time series. 
By making proper changes in the fitness function, a useful algorithm could be obtained for the detection and measurement of periodic signal in time series.

\subsection*{Acknowledgements} 
We thank Prof. Frank Schweitzer for his advice during these investigations. 
We also thank Prof. Hans-Dieter Burkhard for his useful comments and suggestions and Dr. Dagmar Monett for providing us the program \emph{+CARPS (Multiagent  System for Configuring Algorithms in Real Problem  Solving)}.

\bibliographystyle{acs-bibstyle.bst}

\bibliography{/windows/D/PhDDoc/bib/MAS,/windows/D/PhDDoc/bib/StatPhysics,/windows/D/PhDDoc/bib/Math,/windows/D/PhDDoc/bib/Economy,/windows/D/PhDDoc/bib/ML,/windows/D/PhDDoc/bib/ArtificialIntelligence,/windows/D/PhDDoc/bib/Own,/windows/D/PhDDoc/bib/RiskControl,/windows/D/PhDDoc/bib/InvestmentStrategies}

\begin{thebibliography}{40}
\expandafter\ifx\csname natexlab\endcsname\relax\def\natexlab#1{#1}\fi
\expandafter\ifx\csname url\endcsname\relax
  \def\url#1{\texttt{#1}}\fi
\expandafter\ifx\csname urlprefix\endcsname\relax\def\urlprefix{URL }\fi
\expandafter\ifx\csname selectlanguage\endcsname\relax
  \def\selectlanguage#1{\relax}\fi

\bibitem[{Alvarez \emph{et~al.}(2001)Alvarez, Orfila and Tintore}]{alvarez01}
Alvarez, A.; Orfila, A.; Tintore, J. (2001).
\newblock DARWIN: An evolutionary program for nonlinear modeling of chaotic
  time series.
\newblock \emph{Computer Physics Communications} \textbf{136(3)}, 334--349.

\bibitem[{Arrow(1965)}]{Arrow65}
Arrow, J.~K. (1965).
\newblock \emph{Aspects of the Theory of Risk Bearing}.
\newblock Helsinki: Helsinki.

\bibitem[{Artzner \emph{et~al.}(1999)Artzner, Delbaen, Eber and
  Heath}]{Artzner99}
Artzner, P.; Delbaen, F.; Eber, J.-M.; Heath, D. (1999).
\newblock Coherent Measures of Risk.
\newblock \emph{Math. Finance} \textbf{9(3)}, 203--228.

\bibitem[{Bak \emph{et~al.}(1999)Bak, Norrelykke and Shubik}]{Bak-Norrelykke99}
Bak, P.; Norrelykke, S.~F.; Shubik, M. (1999).
\newblock The Dynamics of Money.
\newblock \emph{Physical Review E} \textbf{60(3)}, 2528--2532.

\bibitem[{Branke(1999)}]{Branke99}
Branke, J. (1999).
\newblock Memory Enhanced Evolutionary Algorithms for Changing Optimization
  Problems.
\newblock In: \emph{Proceedings of the Congress on Evolutionary Computation
  CEC'99}. vol.~3, pp. --1882.

\bibitem[{Dawid(1999)}]{Dawid99}
Dawid, H. (1999).
\newblock \emph{Adaptive learning by genetic algorithms: Analytical results and
  applications to economic models}.
\newblock Berlin: Springer, revised second edn.

\bibitem[{Drake and Marks(2002)}]{Drake-Marks02}
Drake, A.~E.; Marks, R.~E. (2002).
\newblock Genetic Algorithms in Economics and Finance: Forecasting Stock Market
  Prices and Foreign Exchange - A review.
\newblock In: S.-H. Chen (ed.), \emph{Genetic Algorithms and Genetic
  Programming in Computational Finance}, Dardrecht: Kluwer Academic. pp.
  29--54.

\bibitem[{Farmer(2001)}]{Farmer01}
Farmer, J.~D. (2001).
\newblock Toward Agent-Based models for Investment.
\newblock In: \emph{Benchmarks and Attibution Analysis}. Association for
  Investment and Management Research, pp. 61--70.

\bibitem[{Farmer \emph{et~al.}(2005)Farmer, Patelli and
  Zovko}]{Farmer-Patelli04}
Farmer, J.~D.; Patelli, P.; Zovko, I.~I. (2005).
\newblock The Predictive Power of Zero Intelligence in Financial Markets.
\newblock In: \emph{Proceedings of the National Academy of Sciences}. vol. 102,
  pp. 2254--2259.

\bibitem[{Forrest(1996)}]{Forrest96}
Forrest, S. (1996).
\newblock Genetic algorithms.
\newblock \emph{ACM Computing Surveys} \textbf{28(1)}, 77--80.
\newblock ISSN 0360-0300.

\bibitem[{Geibel and Wysotzki(2005)}]{Geibel05}
Geibel, P.; Wysotzki, F. (2005).
\newblock Risk-Sensitive Reinforcement Learning Applied to Control under
  Constraints.
\newblock \emph{Journal of Artificial Intelligence Research} \textbf{24},
  81--108.

\bibitem[{Gode and Sunder(1993)}]{Gode-Sunder93}
Gode, D.~K.; Sunder, S. (1993).
\newblock Allocative efficiency of markets with zero-intelligence traders:
  Market as a partial substitute for individual rationality.
\newblock \emph{Journal of Political Economy} \textbf{101}, 119--137.

\bibitem[{Goldberg(1989)}]{Goldberg89}
Goldberg, D.~E. (1989).
\newblock \emph{Genetic {A}lgorithms in {S}earch, {O}ptimization and {M}achine
  {L}earning}.
\newblock Mass., USA: Addison-Wesley.

\bibitem[{Grefenstette(1992)}]{Grefenstette92}
Grefenstette, J.~J. (1992).
\newblock Genetic algorithms for changing environments.
\newblock In: R.~Manner; B.~Manderick (eds.), \emph{Parallel Problem Solving
  from Nature 2.}, Amsterdam: North Holland: Elsevier. pp. 137--144.

\bibitem[{Harvey(1992)}]{Harvey92}
Harvey, I. (1992).
\newblock The SAGE Cross: The mechanics of Recombination for Species with
  Variable-length Genotypes.
\newblock \emph{Parallel Problem Solving from Nature} \textbf{2}, 269--278.
\newblock North-Holland.

\bibitem[{Holland(1975)}]{Holland75}
Holland, J.~H. (1975).
\newblock \emph{Adaptation in Natural and Artificial Systems}.
\newblock Ann Arbor, MI: The University of Michigan Press.

\bibitem[{Jiang and Szeto(2003)}]{Jiang03}
Jiang, R.; Szeto, K.~Y. (2003).
\newblock Extraction of Investment Strategies based on Moving Averages: A
  Genetic Algorithm Approach.
\newblock In: \emph{CIFEr'03}. Hong Kong: IEEE Press, pp. 403--410.

\bibitem[{Kahneman and Tversky(1979)}]{Kahneman-Tversky79}
Kahneman, D.; Tversky, A. (1979).
\newblock Prospect Theory of Decisions under Risk.
\newblock \emph{Econometrica} \textbf{47}, 263--291.

\bibitem[{Kahnemann and Riepe(1998)}]{Kahnemann-Riepe98}
Kahnemann, D.; Riepe, M.~W. (1998).
\newblock Beliefs, preferences and biases investment advisors should know
  about.
\newblock \emph{Journal of Portfolio Management} \textbf{24(4)}.

\bibitem[{Kelly(1956)}]{Kelly56}
Kelly, J.~L. (1956).
\newblock A new Interpretation of Information Rate.
\newblock \emph{The Bell System Technical Journal} .

\bibitem[{Kesten(1973)}]{Kesten73}
Kesten, H. (1973).
\newblock Random difference equations and renewal theory for products of random
  matrices.
\newblock \emph{Acta Math.} \textbf{131}, 207--248.

\bibitem[{LeBaron(2000)}]{LeBaron00}
LeBaron, B. (2000).
\newblock Agent-based computational finance: Suggested readings and early
  research.
\newblock \emph{Journal of Economic Dynamics and Control} \textbf{24},
  679--702.

\bibitem[{Levy and Solomon(1996)}]{Levy-Solomon96}
Levy, M.; Solomon, S. (1996).
\newblock Power laws are logarithmic Boltzmann laws.
\newblock \emph{International Journal of Modern Physics C} \textbf{7},
  595--601.

\bibitem[{Lux and Marchesi(2002)}]{Lux-Marchesi02}
Lux, T.; Marchesi, M. (2002).
\newblock Special issue on heterogeneous interacting agents in financial
  markets.
\newblock \emph{Journal of Economic Behavior and Organization} \textbf{49(2)},
  143--147.

\bibitem[{Magdon-Ismail \emph{et~al.}(2001)Magdon-Ismail, Nicholson and
  Abu-Mostafa}]{Ismail01}
Magdon-Ismail, M.; Nicholson, A.; Abu-Mostafa, Y. (2001).
\newblock Learning in the presence of noise.
\newblock In: S.~Haykin; B.~kosko (eds.), \emph{Intelligent Signal Processing},
  IEEE Press, chap.~3. pp. 120--127.

\bibitem[{Markowitz(1952)}]{markowitz52}
Markowitz, H.~M. (1952).
\newblock Portfolio selection.
\newblock \emph{The Journal of Finance} \textbf{7}, 77--91.

\bibitem[{Marsili \emph{et~al.}(1998)Marsili, Maslov and Zhang}]{marsili98}
Marsili, M.; Maslov, S.; Zhang, Y.-C. (1998).
\newblock Dynamical optimization theory of a diversified portfolio.
\newblock \emph{Physica A} \textbf{253}, 403--418.

\bibitem[{Maslov and Zhang(1998)}]{maslov98}
Maslov, S.; Zhang, Y.-C. (1998).
\newblock Optimal Investment Strategy for Risk Assets.
\newblock \emph{Mathematical Models and Methods in Applied Sciences} .

\bibitem[{Michalewi\c{c}z(1999)}]{Michalewicz99}
Michalewi\c{c}z, Z. (1999).
\newblock \emph{Genetic {A}lgorithms + {D}ata {S}tructures = {E}volution
  {P}rograms}.
\newblock Berlin Heidelberg: Springer, {T}hird, {R}evised and {E}xtended edn.

\bibitem[{Monett(2004)}]{monett04a}
Monett, D. (2004).
\newblock +{CARPS}: {C}onfiguration of {M}etaheuristics {B}ased on
  {C}ooperative {A}gents.
\newblock In: C.~Blum; A.~Roli; M.~Sampels (eds.), \emph{Proceedings of the
  1$^{st}$ International Workshop on Hybrid Metaheuristics, {HM}'2004, at the
  16$^{th}$ European Conference on Artificial Intelligence, {ECAI}'2004}.
  Valencia, Spain, pp. 115--125.

\bibitem[{Navarro and Schweitzer(2003)}]{NavarroSchweitzer03}
Navarro, J.~E.; Schweitzer, F. (2003).
\newblock The Investors Game: A Model for Coalition Formation.
\newblock In: L.~Czaja (ed.), \emph{Proceedings of the Workshop on Concurrency,
  Specification \textrm{\&} Programming, {CS \textrm{\&} P}'2003}. Czarna,
  Poland: Warsaw University, vol.~2, pp. 369--381.

\bibitem[{Navarro-Barrientos \emph{et~al.}(2007)Navarro-Barrientos,
  Cantero-Alvar\'ez, Rodriguez and Schweitzer}]{NavarroPhysicaA}
Navarro-Barrientos, J.~E.; Cantero-Alvar\'ez, R.; Rodriguez, J. F.~M.;
  Schweitzer, F. (2007).
\newblock Investments in random environments.
\newblock Physica A (2007), doi:10.1016/j.physa.2007.11.029.

\bibitem[{Neely \emph{et~al.}(1997)Neely, Weller and Dittmar}]{Neely97}
Neely, C.; Weller, P.; Dittmar, R. (1997).
\newblock Is Technical Analysis in the Foreign Exchange Market Profitable? A
  Genetic Programming Approach.
\newblock \emph{The Journal of Financial and Quantitative Analysis}
  \textbf{32(4)}, 405--426.

\bibitem[{Redner(1990)}]{Redner90}
Redner, S. (1990).
\newblock Random multiplicative processes: An elementary tutorial.
\newblock \emph{American Journal of Physics} \textbf{58}, 267--273.

\bibitem[{Sankoff and Kruskall(1983)}]{sankoff83}
Sankoff, D.; Kruskall, J.~B. (1983).
\newblock \emph{Time Warps, String Edits and Macromolecules: The Theory and
  Practice of Sequence Comparison}.
\newblock Addison-Wesley.

\bibitem[{Schulenburg and Ross(1999)}]{schulenburg99evolutionary}
Schulenburg, S.; Ross, P. (1999).
\newblock An evolutionary approach to modelling the behaviours of financial
  traders.
\newblock In: S.~Brave; A.~S. Wu (eds.), \emph{Late Breaking Papers at the 1999
  Genetic and Evolutionary Computation Conference}. Orlando, Florida, USA, pp.
  245--253.
\newblock \urlprefix\url{citeseer.ist.psu.edu/schulenburg99evolutionary.html}.

\bibitem[{Schulenburg and Ross(2001)}]{schulenburg01strength}
Schulenburg, S.; Ross, P. (2001).
\newblock Strength and Money: An {LCS} Approach to Increasing Returns.
\newblock \emph{Lecture Notes in Computer Science} \textbf{1996}, 114--137.
\newblock \urlprefix\url{citeseer.ist.psu.edu/schulenburg01strength.html}.

\bibitem[{Sornette and Cont(1997)}]{Sornette-Cont97}
Sornette, D.; Cont, R. (1997).
\newblock Convergent Multiplicative Processes Repelled from Zero: Power Laws
  and Truncated Power Laws.
\newblock \emph{Journal of Physics} \textbf{1(7)}, 431.

\bibitem[{Szpiro(1997)}]{szpiro97PhysRevE.55.2557}
Szpiro, G.~G. (1997).
\newblock Forecasting chaotic time series with genetic algorithms.
\newblock \emph{Phys. Rev. E} \textbf{55(3)}, 2557--2568.

\bibitem[{Turing(1960)}]{turing60}
Turing, G. (1960).
\newblock An introduction to matched filters.
\newblock \emph{IEEE Transactions on Information Theory} .

\end{thebibliography}

\end{document}